\documentclass[aps, reprint, prb, twocolumn, amsmath, amssymb, groupedaddress, 10pt,
citeautoscript, longbibliography]{revtex4-1}

\usepackage{epsfig}
\usepackage{subfigure}
\usepackage{graphicx}
\usepackage{amsfonts}
\usepackage[figuresright]{rotating}
\usepackage{amssymb}
\usepackage{amsmath}
\usepackage{psfrag}
\usepackage{esint} 
\usepackage{dcolumn}
\usepackage{hyperref}
\hypersetup{colorlinks,allcolors=blue} 

\usepackage{multirow}
\usepackage{siunitx}
\usepackage{xcolor}
\usepackage{lipsum}
\usepackage{bm}
\usepackage{soul}

\usepackage{mathrsfs}
\usepackage{chemformula}

\global\hyphenpenalty=100000

\begin{document}

\title{Spontaneous $\pi$ flux trapping in granular rings of unconventional superconductors} 

\author{Junyi Zhang}
\affiliation{Department of Physics and Astronomy, Johns Hopkins University, Baltimore, Maryland 21218, USA}
\author{Yi Li}
\affiliation{Department of Physics and Astronomy, Johns Hopkins University, Baltimore, Maryland 21218, USA}

\date{\today}

\begin{abstract}
We study Josephson couplings in unconventional superconductors and generalize the Sigrist-Rice formula by incorporating symmetry constraints and interface orientation disorder. Applying this framework to granular superconducting rings, we establish a no-go result that single-band chiral superconductors cannot spontaneously trap a magnetic flux. This rules out chiral $p$-wave pairing in $\beta$-Bi$_2$Pd, in light of the half-quantum flux observed in recent Little-Parks experiments. Incorporating the full crystalline and time-reversal symmetries, we show that a fully gapped helical equal-spin pairing state, naturally stabilized by spin-orbit coupling arising from local inversion-symmetry breaking, is instead favored. We further find that granular rings of such superconductors can trap a spontaneous $\pi$ flux in a manner robust against interface disorder.
\end{abstract}

\maketitle

Supercurrents in a Josephson junction arise from coherent quantum tunneling~\cite{Josephson1962,Anderson1963}. 
In a superconducting quantum interference device (SQUID), 
the supercurrents through two junctions interfere, analogous to optical double-slit interference, 
with the modulation controlled by an external magnetic flux~\cite{Jaklevic1964}. 
When tunneling occurs through a ferromagnetic barrier or a region containing correlated magnetic impurities~\cite{Shiba1969,Bulaevskii1977,Buzdin1982,Glazman1989,Spivak1991}, the Josephson coupling in conventional superconductors can change sign, forming a so-called $\pi$ junction~\cite{Bulaevskii1977}.
A superconducting ring incorporating such a $\pi$ junction becomes frustrated and supports a spontaneous circulating supercurrent in its ground state, 
trapping a magnetic flux of half the superconducting flux quantum ($\pi$ flux) even in the absence of an external magneticfield~\cite{Bulaevskii1977}.

The phase sensitivity of Josephson junction also provides a powerful tool for probing the internal structure of unconventional superconducting order parameters.
The approach was first proposed theoretically by Geshkenbein, Larkin, and Barone~\cite{Geshkenbein1986,Geshkenbein1987}, 
and later extended to high-$T_c$ cuprates by Sigrist and Rice~\cite{Sigrist1992,Sigrist1995}.
Experimentally, phase-sensitive Josephson interferometry in corner junction configurations 
by Van Harlingen and collaborators identified $d_{x^2 - y^2}$-wave pairing in \ch{YBa_2Cu_3O_{7-$x$}} (YBCO)~\cite{Wollman1993, Wollman1995, VanHarlingen1995},
and $E_{2u}$ chiral $f$-wave pairing order symmetry~\cite{Strand2009} in the heavy-fermion superconductor \ch{UPt_3}~\cite{Stewart1984, Aeppli1988, Aeppli1989, Fisher1989, Sauls1994, Tou1996,  Park1996, Joynt2002}.
A landmark phase-sensitive demonstration of $d$-wave pairing symmetry was provided by the tricrystal ring experiments of Tsuei, Kirtley, and collaborators~\cite{Tsuei1994, Kirtley1996, Tsuei2000}. 
They fabricated rings from YBCO and $\ch{Bi_2Sr_2CaCu_2O_{8+$\delta$}}$ with three grains of specific crystallographic orientations, 
and realized frustrated and unfrustrated Josephson loops. 
Scanning SQUID microscopy revealed a spontaneously trapped $\pi$ flux at the center of the frustrated rings, 
but not in the unfrustrated controls, 
providing direct evidence for the sign-changing order parameter characteristic of $d_{x^2 - y^2}$-wave pairing in cuprates~\cite{Tsuei2000}.

The Little–Parks effect~\cite{Little1962}, 
the oscillatory modulation of the transition temperature or critical current with external magnetic flux through a superconducting ring, 
also serves as a phase-sensitive probe of the pairing symmetry.
Recent observations of anomalous Little–Parks oscillations in unconventional superconductors,
including \ch{CsV3Sb5}~\cite{Ge2024, Zhang2024, Pan2024}, 4Hb-\ch{TaS2}~\cite{Almoalem2024, Fischer2023b}, and $\beta$-\ch{Bi2Pd}~\cite{Li2019, Xu2024, Li2024}, motivate the present theoretical study, 
which focuses on resolving the puzzles in $\beta$-\ch{Bi_2Pd}.
$\beta$-\ch{Bi_2Pd} consists of centrosymmetric \ch{Bi}-\ch{Pd}-\ch{Bi} trilayers stacked in a body-centered tetragonal structure with space group $I4/mmm$~\cite{Herrera2015}.
Its superconducting state is characterized by a nodeless $s$-wave-like gap and the absence of time-reversal symmetry breaking, 
as established by multiple spectroscopic measurements~\cite{Herrera2015, Lv2017, Kacmarcik2016, Chen2020, Biswas2016}.
Nevertheless, the observation of an anomalous $\pi$ phase shift in polycrystalline rings, 
but absent in epitaxial ones, 
in Little–Parks measurements suggests the presence of nontrivial Josephson couplings between grains~\cite{Li2019}. 
Moreover, in Geshkenbein–Larkin–Barone–type composite rings composed of epitaxial $\beta$-\ch{Bi2Pd} and \ch{Nb}, 
a $\pi$ phase shift appears only for antiparallel, 
but not parallel, 
junction interface orientations, 
providing more direct evidence for an anisotropic odd-parity pairing component~\cite{Xu2024}.

To clarify how the anomalous Little–Parks oscillations and the underlying symmetries constrain the superconducting pairing in $\beta$-\ch{Bi2Pd}, 
we analyze spontaneous flux trapping in granular superconducting rings with anomalous Josephson couplings of unconventional pairing states.
We establish a no-go result that 
granular rings composed of single-band chiral superconductors \emph{cannot} support spontaneous flux trapping, 
regardless of the interface or grain orientations.
In contrast, we find that a time-reversal invariant helical equal-spin (HES) pairing  state can indeed support spontaneous $\pi$-flux trapping.
Unlike in real $d$-wave cuprates where the effect is sensitive to interface disorder~\cite{Tsuei2000},
the spontaneous flux trapping induced by HES pairing remains robust against interface orientation disorder.
Moreover, 
by considering the full space-group symmetry, 
we show that spin–orbit coupling induced by local inversion-symmetry breaking~\cite{Fischer2023} generates a ``hidden spin polarization'' on the Fermi surfaces, 
which naturally favors the HES pairing state.
These results clarify the nature of the pairing with puzzling features observed in spectroscopic and transport experiments.


\begin{figure}[htb]
\includegraphics[width=\linewidth]{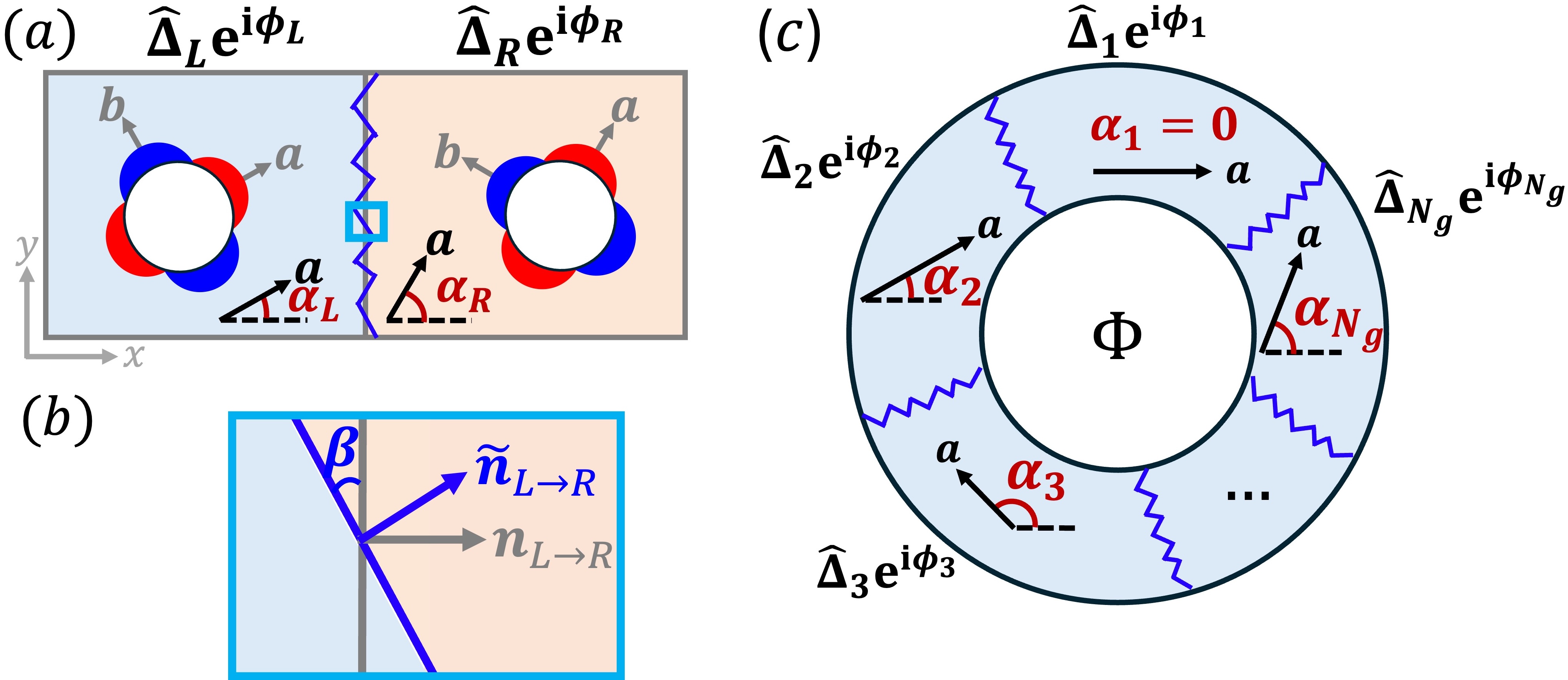}
\caption{\label{fig:JJ}
(a) Josephson junction between two superconducting grains. The pairing order parameters of the grains are $\hat{\Delta}_{L,R}\mathrm{e}^{\mathrm{i} \phi_{L,R}}$ on the left and right, respectively. 
The crystallographic $a$-axis of each grain is rotated by an angle $\alpha_{L(R)}$ relative to the normal direction of the junction interface $\mathbf{n}_{L\rightarrow R}$. 
(b) A zoomed-in view of a local segment of the junction interface, highlighted by the blue box in (a).
In the presence of interface orientation disorder, the local normal direction $\tilde{\mathbf{n}}_{L \rightarrow R}$ (blue) deviates from the average normal direction $\mathbf{n}_{L \rightarrow R}$ (gray) by an angle $\beta$.
(c) A superconducting ring composed of $N_g$ grains, enclosing a magnetic flux $\Phi_B$ through its central hole.
The $n$-th grain has its crystallographic $a$-axis oriented at an angle $\alpha_n$ with respect to the $x$-axis,
and its pairing order parameter is $\hat{\Delta}_n \mathrm{e}^{\mathrm{i} \phi_n}$.
Without loss of generality, we set $\alpha_1 = 0$ in the first grain.
}
\end{figure}

We determine the first-order Josephson couplings based on symmetry considerations for both clean and dirty junctions~\footnote{Following Ref.~\citenum{Tsuei2000}, 
``dirty'' junctions refer to those with interface orientation disorder.}.
Applied to $d$-wave cuprates, 
our formalism generalizes the Sigrist–Rice formula~\cite{Sigrist1992,Sigrist1995} in the clean limit, 
adding an additional symmetry-allowed term previously derived via series expansion by Walker and Luettmer-Strathmann~\cite{WLS1996}, 
here obtained directly from symmetry arguments.
In the dirty limit, 
the result unifies the earlier form suggested by Tsuei and collaborators~\cite{Tsuei1994,Tsuei2000}
while providing a systematic and physically transparent framework.

Let us first consider two superconducting grains in contact, 
forming a Josephson junction as illustrated in Fig.~\ref{fig:JJ} (a).
Each grain hosts a uniform pairing order $\hat{\Delta}_n \mathrm{e}^{\mathrm{i} \phi_n}$, $n = L, R$ in its bulk,
where $\hat{\Delta}_n$ is the pairing function defined relative to the crystallographic orientation of the grain,
and $\phi_n$ is the $U(1)$ phase remaining arbitrary for an isolated grain due to spontaneous symmetry breaking in the superconducting state.
When the two superconducting grains are weakly coupled, 
a Josephson supercurrent can tunnel through the interface at zero voltage bias.
The first-order Josephson coupling is described by a free energy of the form
\begin{equation}
\label{eq:JC_JCFreeEnergy_Sesquilinear}
\begin{split}
F_J = \mathcal{I}[\hat{\Delta}_L,
\hat{\Delta}_R]\mathrm{e}^{\mathrm{i}(\phi_L - \phi_R)} + \text{c.c.},
\end{split}
\end{equation}
where $\mathcal{I}$ denotes the Josephson coupling amplitude,
which acts as a sesquilinear functional of the pairing functions, 
linear in $\hat{\Delta}_L$
and antilinear in $\hat{\Delta}_R$.
It is important to note that 
the Josephson free energy $F_J$ is gauge invariant,
depending only on the phase difference $(\phi_L - \phi_R)$, 
while each superconducting grain can spontaneously adjust its $U(1)$ phase to minimize the free energy.
For specific microscopic models, 
$\mathcal{I}$ can be derived using linear response theory,
generalizing the approach of Ambegaokar and Baratoff~\cite{Ambegaokar1963, Ambegaokar1963Erratum}
to unconventional multiband superconductors (see Supplemental Material  Sec.~\ref{sec:SI_AB} for details).
More generally, we expand the pairing functions $\hat{\Delta}_n$ of each grain
in terms of its basis set $\mathcal{B}_n = \{\hat{\Delta}^{n}_{b}\}$ as
$
\hat{\Delta}_n = \sum_{b \in \mathcal{B}_n} d^{n}_{b}\hat{\Delta}^{n}_{b},
$
where the complex coefficients $d^{n}_{b}$ specify the amplitudes in the corresponding pairing channels.
Because $\mathcal{I}$ is sesquilinear, 
the Josephson coupling amplitude takes the form
$
\mathcal{I}[\hat{\Delta}_L, \hat{\Delta}_R]
= \sum_{
b_1 \in \mathcal{B}_L, 
b_2 \in \mathcal{B}_R} d^{L}_{b_1} (d_{b_2}^{R})^* I_{b_1,b_2},
$
where $I_{b_1,b_2} = \mathcal{I}[\hat{\Delta}^{L}_{b_1}, \hat{\Delta}^{R}_{b_2}]$
are the matrix elements of $\mathcal{I}$ in this basis.

The symmetries of the junction impose stringent constraints on the functional $\mathcal{I}$.
Consider a lateral Josephson junction, 
as illustrated in Fig.~\ref{fig:JJ} (a), 
with the $x$-axis normal to the interface.
The mirror reflection
$\mathcal{M}_y: (x, y) \mapsto (x, -y)$,
is a symmetry of this junction geometry
and allows the basis pairing functions to be classified by their mirror eigenvalues,
i.e.,
$\mathcal{M}_y \hat{\Delta}^n_{b} = m_b \hat{\Delta}^n_{b}$, 
with $m_b=\pm 1$.
Mirror symmetry immediately forbids Josephson couplings between channels of opposite mirror parity.
In addition,
the representations of the crystalline rotational symmetry determine 
how the Josephson coupling amplitude depends on the relative grain orientations
thereby generalizing the Sigrist–Rice formula beyond $d$-wave cuprates.
If the crystallographic axes of grain $n$ are rotated by an angle $\alpha_n$ [Fig.~\ref{fig:JJ} (a)], 
its pairing function transforms as
$
\hat{\Delta}_n(\alpha_n) 
= \sum_{b',b \in \mathcal{B}_n} d^n_{b'} R^n_{b',b}(\alpha_n)\hat{\Delta}_{b}^{n},
$
where $R^n_{b',b}(\alpha_n)$ is the matrix representation of the rotation in the basis $\mathcal{B}_n$.
Using the sesquilinearity of $\mathcal{I}$,
the Josephson coupling amplitude between two rotated grains takes the general form
\begin{equation}\label{eq:JC_AngularFormFactor}
\begin{split}
\mathcal{I}[\hat{\Delta}_L^{\alpha_L} , \hat{\Delta}_R^{\alpha_R} ]
=&\sum_{b'_1,b_1, b'_2,b_2 } 
d_{b'_1}^L R^L_{b'_1,b_1}({\alpha_L}) \\
& \times (d_{b'_2}^R)^* [R^R_{b'_2,b_2}({\alpha_R})]^*I_{b_1,b_2}.
\end{split}
\end{equation}
Taken together, Eqs.~\eqref{eq:JC_JCFreeEnergy_Sesquilinear} and~\eqref{eq:JC_AngularFormFactor} provide a systematic, 
symmetry-based method for determining the first-order Josephson coupling between arbitrarily oriented grains.

As a benchmark, we apply our formalism to single-band $d$-wave superconductors.
In the basis $\mathcal{B}_d = \{ \hat{\Delta}_{d_{x^2-y^2}},\hat{\Delta}_{d_{xy}} \}$, 
which spans the relevant two-dimensional $d$-wave representation and forms eigenstates of the mirror reflection $\mathcal{M}_y$,
only intra-channel couplings survive,
$I_{d_{x^2-y^2},d_{x^2-y^2}} = -I_v$ and $I_{d_{xy},d_{xy}} = -I_d$. 
This yields the generalized Josephson coupling
$
\mathcal{I}[\hat{\Delta}_{d}(\alpha_L), \hat{\Delta}_{d}(\alpha_R)]
= - I_{v} \cos(2\alpha_L) \cos(2\alpha_R) - I_{d} \sin(2\alpha_L) \sin(2\alpha_R)
$
[Supplemental Material Sec.~\ref{sec:SI_JCpdSC} Eq.~\eqref{eq:SI_JCpdSC_dWaveJCAngularDep}]. 
This result, 
first derived by Walker and Luettmer-Strathmann~\cite{WLS1996} 
using a series-expansion approach, 
generalizes the original Sigrist–Rice formula~\cite{Sigrist1992, Sigrist1995}, that only includes the first term ($\propto I_v$).
The second term, 
although symmetry-allowed, 
has been largely overlooked in analyses of tricrystal experiments~\cite{Tsuei1994, Tsuei2000}.
Microscopic studies by Tanaka and collaborators showed that 
zero-energy Andreev bound states in real $d$-wave superconductors can further enhance the Josephson coupling beyond the Sigrist–Rice prediction~\cite{Tanaka1996, Tanaka1997, Kashiwaya2000, Tanaka2024}.
In this work, we focus on candidate pairings for $\beta$-\ch{Bi_2Pd} that may support chiral or helical surface states.
Because the zero-energy surface states appear only at isolated time-reversal invariant momenta, 
the corrections of zero-energy bound states are negligible, allowing our symmetry-based analysis to capture the essential physics.

Interface disorder can significantly modify the angular dependence of Josephson couplings. 
We model it as random local fluctuations of the interface orientation as schematically shown in Fig.~\ref{fig:JJ} (b), 
where the local normal direction $\tilde{\mathbf{n}}_{L\rightarrow R}$ deviates from the average junction normal direction $\mathbf{n}_{L\rightarrow R}$ 
by an angle $\beta(y)$, 
yielding effective local grain angles $\tilde{\alpha}_n(y) = \alpha_n - \beta(y)$. 
The total Josephson coupling is obtained by summing the local contributions,
$
\bar{\mathcal{I}} = ({1}/{W_J}) \int \mathrm{d}y \mathcal{I}[\hat{\Delta}^{\tilde{\alpha}_L(y)} (\mathbf{k}_1), \hat{\Delta}^{\tilde{\alpha}_R(y)} (\mathbf{k}_2)],
$
where $W_J$ is the junction width. 
When $W_J$ greatly exceeds the correlation length of the disorder, 
the integral can be effectively evaluated as an ensemble average over $\beta$,
$
\bar{\mathcal{I}} = \overline{\mathcal{I}[\hat{\Delta}^{\tilde{\alpha}_L} (\mathbf{k}_1), \hat{\Delta}^{\tilde{\alpha}_R} (\mathbf{k}_2)]}_\beta.
$
Applied to real $d$-wave superconductors, 
the disorder-averaged generalized Sigrist–Rice formula contains two angular terms proportional to $\cos(2\alpha_L \pm 2\alpha_R)$ [see Eq.~\eqref{eq:SI_JCpdDisorder_AngularFormFactorDisorderAvgdWave}].
In the dirty limit, the term involving the angle sum is exponentially suppressed, 
reducing the expression to the form used by Tsuei and collaborators~\cite{Tsuei1994,Tsuei2000}.
This provides a physical justification that does not rely on the fine-tuned condition $I_v = -I_d$~\cite{WLS1996,Tsuei2000}.
For real $p$-wave pairing, however, the angle sum contribution, though still suppressed, 
can remain as large as $\sim 30\%$ of the angle difference term and is therefore not negligible
(see Supplemental Material Sec.~\ref{sec:SI_Disorder}).
It is also worth noting that Josephson couplings that depend only on the angle difference $(\alpha_L - \alpha_R)$ 
retain their form even in the presence of interface orientation disorder.

Next, we consider spontaneous flux trapping in a granular superconducting ring composed of $N_g$ grains, 
as schematically shown in Fig.~\ref{fig:JJ} (c), 
where Josephson junctions form at grain boundaries and a magnetic flux $\Phi_B$ threads the central hole.
The total free energy has three contributions
$F = F_S + F_J + F_L$, 
where
$F_S = \sum_{n}F_S^{(n)}$ is the sum of the bulk condensation energies of the grains,
$F_J= \sum_{n} F_J^{(n,n+1)}$ is the Josephson coupling energy at the interfaces,
and $F_L$ is the self-inductance energy associated with the circulating supercurrent.
We adopt two simplifying but physically well-motivated assumptions.
First, the grains are assumed to be larger than the superconducting coherence length, 
ensuring that the pairing function remains fixed within each grain.
Second, for the mesoscopic devices of Refs.~\citenum{Li2019} and~\citenum{Xu2024}, 
the self-inductance energy $F_L$ is negligible compared to $F_S$ and $F_J$.
Under these conditions, the low-energy physics is governed entirely by the minimization of the Josephson energy $F_J$ with respect to the $U(1)$ phase degrees of freedom.  

In the presence of a magnetic flux, 
the phases in Eq.~\eqref{eq:JC_JCFreeEnergy_Sesquilinear} 
are modified as
$\phi_n = \tilde{\phi}_n + \Phi_n$,
where $\tilde{\phi}_n$ is the uniform $U(1)$ phase of grain $n$,
and $\Phi_n (\mathbf{r}) = (e^*/\hbar)\int^{\mathbf{r}} \mathbf{A}\cdot \mathrm{d}\mathbf{r}$
is the magnetic phase shift induced by the vector-potential.
The Josephson coupling energy then becomes
\begin{equation}\label{eq:FreeEnergy_SuperconductingRing}
\begin{split}
F_J =& \sum_{n=1}^{N_g} -I_{n,n+1} \cos\Big[(\tilde{\phi}_{n}-\tilde{\phi}_{n+1}) \\ 
& + \delta \varphi_{n,n+1}
+ (\Phi_{n,+} - \Phi_{n+1,-})\Big],
\end{split}
\end{equation}
where
the intrinsic phase shift $\delta \varphi_{n,n+1}$ 
and the coupling amplitude $-I_{n,n+1}$  follow from Eq.~\eqref{eq:JC_AngularFormFactor},
and
$\Phi_{n,\pm}$ denotes the magnetic phase shift from the vector potential at the interfaces between grain $n$ 
and its neighbors $n\pm1$.
Without loss of generality, 
we take $I_{n,n+1} \geq 0$.
If $I_{n,n+1}<0$ 
the sign can be absorbed by
$I_{n,n+1}\rightarrow |I_{n,n+1}|$ 
and  $\delta\varphi_{n,n+1} \rightarrow \delta\varphi_{n,n+1}+\pi$.
Minimizing $F_J$ with respect to the $U(1)$ phases
yields the conditions of the phases
$ (\tilde{\phi}_{n}-\tilde{\phi}_{n+1}) 
+ \delta \varphi_{n,n+1}
+ (\Phi_{n,+} - \Phi_{n+1,-}) = 2\pi k_n$,
where $k_n \in \mathbb{Z}$ for $n = 1,2,\dots, N_g$.
Since the accumulated magnetic phase around the ring satisfies
$\Phi = \sum_{n} (\Phi_{n,+} - \Phi_{n+1,-}) = 2\pi \Phi_B/\phi_0$,
summing the phase constraints over all junctions yields the criterion for nontrivial flux trapping,
\begin{equation}\label{eq:FreeEnergy_FluxTrappingCriterion}
\begin{split}
\Phi = 2\pi K -\sum_{n}\delta \varphi_{n,n+1} \not\equiv 0 \pmod{2\pi}
\end{split}
\end{equation}
where $K=\sum_n k_n \in \mathbb{Z}$.
In particular, trapping a $\pi$ flux requires 
$\sum_n \delta \varphi_{n,n+1} \equiv \pi \pmod{2\pi}$.
In Supplemental Material Sec.~\ref{sec:SI_3GR}, 
we demonstrate that Eq.~\eqref{eq:FreeEnergy_FluxTrappingCriterion}, 
when evaluated with the Sigrist–Rice Josephson couplings, reproduces the phase diagrams of three-grain cuprate rings reported by Tsuei and collaborators~\cite{Tsuei2000}.

An intriguing finding of this work is a no-go result that 
granular rings made of single-band chiral superconductors with basal-plane rotational symmetry cannot spontaneously trap magnetic flux. 
For a chiral superconductor whose pairing function transforms under a rotation $R_z(\alpha)$ as
$
\hat{\Delta}_{m}^{\alpha}(\mathbf{k})
= \hat{\Delta}_{m} \mathrm{e}^{-\mathrm{i} m \alpha}
$
with chiral quantum number $m$
[the eigenvalue of the $z$-component of the Cooper-pair total angular momentum, e.g., $m=\pm 2$ for ($d_{x^2-y^2} \pm \mathrm{i}d_{xy})$-pairing],
the Josephson coupling free energy takes the form
\begin{equation}\label{eq:ChiralSC_ChiraldWaveFreeEnergy}
\begin{split}
F_J =& \sum_{n} -I_{n,n+1} \cos\Big[(\tilde{\phi}_{n}-\tilde{\phi}_{n+1})
\\
&- m(\alpha_{n}-\alpha_{n+1})
+ (\Phi_{n,+}-\Phi_{n+1,-})\Big],
\end{split}
\end{equation}
as derived in Supplemental Material Sec.~\ref{sec:SI_JCpdSC}.
The corresponding intrinsic phase shift is simply
$\delta \varphi_{n,n+1} = -m(\alpha_n-\alpha_{n+1})$,
depending only on the relative orientations of neighboring grains.
Since the sum of all orientation differences around any closed ring necessarily vanishes,
$\sum_n \delta\varphi_{n,n+1} = 0$,
the criterion for spontaneous flux trapping, Eq.~\eqref{eq:FreeEnergy_FluxTrappingCriterion}, 
immediately implies that no spontaneous magnetic flux can be trapped, 
regardless of the number of grains, their shapes, or their arrangement.
This no-go result applies equally to chiral $p$- and chiral $d$-wave states.
More generally, any rotationally symmetric pairing state with a fixed chiral quantum number $m$ produces a trivial total phase winding, and consequently granular chiral superconductors cannot exhibit spontaneous flux trapping in ring geometries.

The observation of a $\pi$ phase shift in Geshkenbein–Larkin–Barone type interferometry suggests odd-parity pairing in 
$\beta$-\ch{Bi2Pd}~\cite{Xu2024}, 
and spectroscopic measurements indicate a nodeless gap~\cite{Herrera2015,Lv2017,Kacmarcik2016,Chen2020,Biswas2016,Xu2024}. 
These features make chiral $p$-wave pairing an initially appealing candidate.
However, the half-quantum flux phase shift observed in the Little–Parks experiments on polycrystalline $\beta$-\ch{Bi2Pd} rings~\cite{Li2019},
together with the no-go result established above, 
rules out the single-band chiral $p$-wave pairing
that cannot support spontaneous $\pi$ flux in a granular ring. 
This motivates the search for alternative pairing symmetries consistent with all these observations.

Although $\beta$-\ch{Bi2Pd} is centrosymmetric (space group $I4/mmm$),
its Fermi surfaces,
mainly derived from \ch{Bi} $p$-orbitals 
forming quasi-two-dimensional bands,
can experience spin–orbit coupling 
arising from local inversion-symmetry breaking~\cite{Fischer2023}. 
This essential feature has been largely overlooked in previous discussions of its superconducting state.
To incorporate the full crystallographic symmetry, 
we consider a minimal model on a bilayer square lattice in $A$-$A$ stacking, 
as schematically shown in Fig.~\ref{fig:SFig_CrystalStructure}. 
Therefore, the band Hamiltonian has the form
\begin{equation}\label{eq:HESP_BandHam}
\begin{split}
\mathcal{H}_b =& \sum_{\mathbf{k}}
\Big\{
\sum_{l,s}
[\epsilon_0(\mathbf{k})-\mu] c_{\mathbf{k},l,s}^\dagger c_{\mathbf{k},l,s}\\
&+ 
\sum_{l}\sum_{s,s'}
\mathbf{g}_l(\mathbf{k}) \cdot \bm{\sigma}_{s,s'}
c_{\mathbf{k},l,s}^\dagger c_{\mathbf{k},l,s'}\\
&+
\sum_{s}
(t_{\perp})
c_{\mathbf{k},l=1,s}^\dagger c_{\mathbf{k},l=2,s} + \text{h.c.}
\Big\},
\end{split}
\end{equation}
where $l$ is the layer index, and $s,s'$ are the spin indices.
The first term in Eq.~\eqref{eq:HESP_BandHam} describes the in-plane kinetic energy, 
the second introduces the layer-resolved spin–orbit coupling, 
and the last term accounts for interlayer hopping (see Supplemental Material Sec.~\ref{sec:SI_M2BM} for details).
Although the crystal as a whole is inversion symmetric, 
the inversion center lies between the two \ch{Bi} layers,
therefore, each individual layer lacks mirror reflection symmetry $\mathcal{M}_z$
This local inversion asymmetry admits a spin-orbit coupling of Rashba form with 
$\mathbf{g}_{l}(\mathbf{k}) \propto \mathbf{e}_z \times \mathbf{k}$.
Global inversion symmetry requires that the polar fields of the two layers point in opposite directions, 
which enforces $\mathbf{g}_{+}(\mathbf{k}) = \mathbf{g}_{-}(-\mathbf{k})$,
so that the full bilayer system remains inversion symmetric even though each layer individually is not.

The Fermi surfaces remain doubly degenerate due to the global inversion and time-reversal symmetries, 
yet they exhibit ``hidden spin polarization''~\cite{Fischer2023},
i.e., a helicity with spin–momentum locking,
arising from spin–orbit coupling generated by the local inversion symmetry breaking.
If the superconductivity originates from a conventional short-range 
attractive interaction mediated by phonons~\cite{Zheng2017,Saib2017},
the effective pairing on the Fermi surfaces can be obtained by 
projecting the pairing interaction onto them,
which leads to the effective gap function
\begin{equation}\label{eq:HESP_SCProjFS14}
\begin{split}
\hat{\Delta}_{ES}(\mathbf{k})
=
\Delta_0 \begin{pmatrix}
+\mathrm{i} \mathrm{e}^{-\mathrm{i} \theta_{\mathbf{k}}} & 0\\
0& -\mathrm{i} \mathrm{e}^{+\mathrm{i} \theta_{\mathbf{k}}} 
\end{pmatrix}.
\end{split}
\end{equation}
See Supplemental Materials Sec.~\ref{sec:SI_M2BM} for detailed derivations.
This corresponds to an equal-spin pairing state,
in which the pairing functions of the two effective spin components carry opposite chiralities 
and are related by time-reversal symmetry.
In this way, the spin–momentum locking on the Fermi surfaces induced by spin-orbit coupling
naturally stabilizes this HES pairing.

\begin{figure}[tb]
\includegraphics[width=0.85\linewidth]{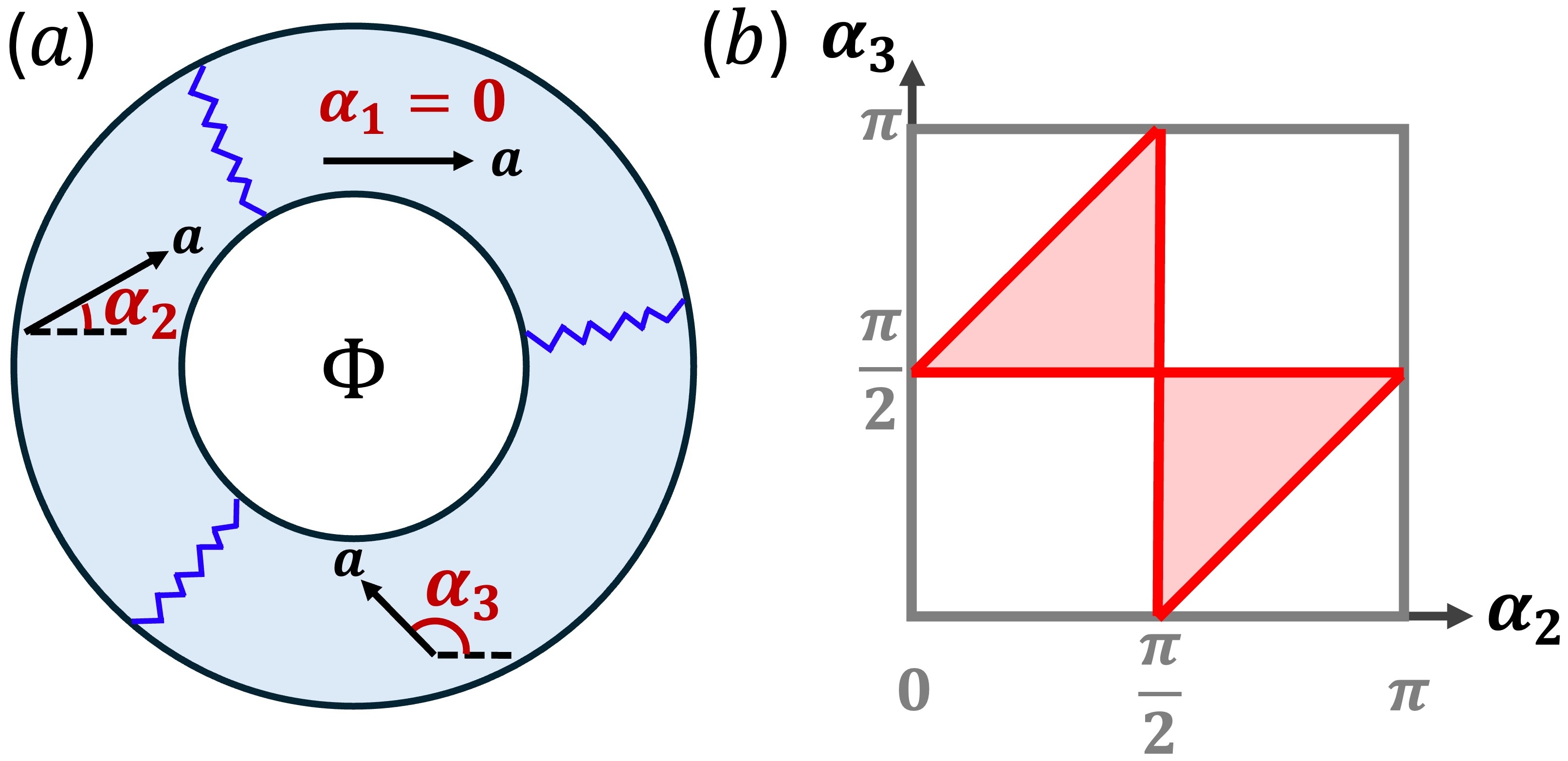}
\caption{\label{fig:3GR} 
Three-grain superconducting ring.
(a) Geometric configuration of the ring, with grain orientations labeled by 
$\alpha_{1,2,3}$.
(b) Phase diagram for three-grain rings of HESP superconductor plotted in the $(\alpha_2,\alpha_3)$ plane with $\alpha_{1}=0$. 
The red-shaded region indicates configurations supporting spontaneous 
$\pi$ flux trapping.
}
\end{figure}

The Josephson free energy between two HES pairing superconductors receives contributions from both effective spin sectors,
$F_J = F_J^{\uparrow} + F_J^{\downarrow}$,
where each sector contributes $F_J^{\sigma} = -I^{\sigma} \cos[(\phi_L-\phi_R)+ m_{\sigma}(\alpha_L-\alpha_R)]$
with chirality $m_{\sigma} = \pm 1$.
The time-reversal symmetry enforces $I^{\uparrow} = I^{\downarrow} = I$
Consequently, the total Josephson free energy of a superconducting ring composed of $N_g$ grains takes the form
\begin{equation}\label{eq:HESP_HESPFreeEnergy}
\begin{split}
F_J =&\sum_{n=1}^{N_g} 
-2I_{n,n+1}\cos (\alpha_{n} - \alpha_{n+1})\\
&\times
\cos\Big[(\tilde{\phi}_{n}-\tilde{\phi}_{n+1}) 
+ (\Phi_{n,+} -\Phi_{n+1,-}) \Big].
\end{split}
\end{equation}
In contrast to Eq.~\eqref{eq:ChiralSC_ChiraldWaveFreeEnergy},
the Josephson coupling of HES superconductors acquires an angular dependence 
only through a multiplicative cosine factor.
This form is analogous to that of single-band $d$-wave superconductors in the dirty junction limit
[Supplemental Material Sec.~\ref{sec:SI_Disorder},
Eq.~\eqref{eq:SI_JCpdDisorder_GeneralizedSRForumlaDisorder}].
As a result, granular rings of HES superconductors can spontaneously trap a 
$\pi$ flux for certain grain configurations, and the flux trapping remains robust against interface-orientation disorder.

\begin{figure}[tb]
\includegraphics[width=\linewidth]{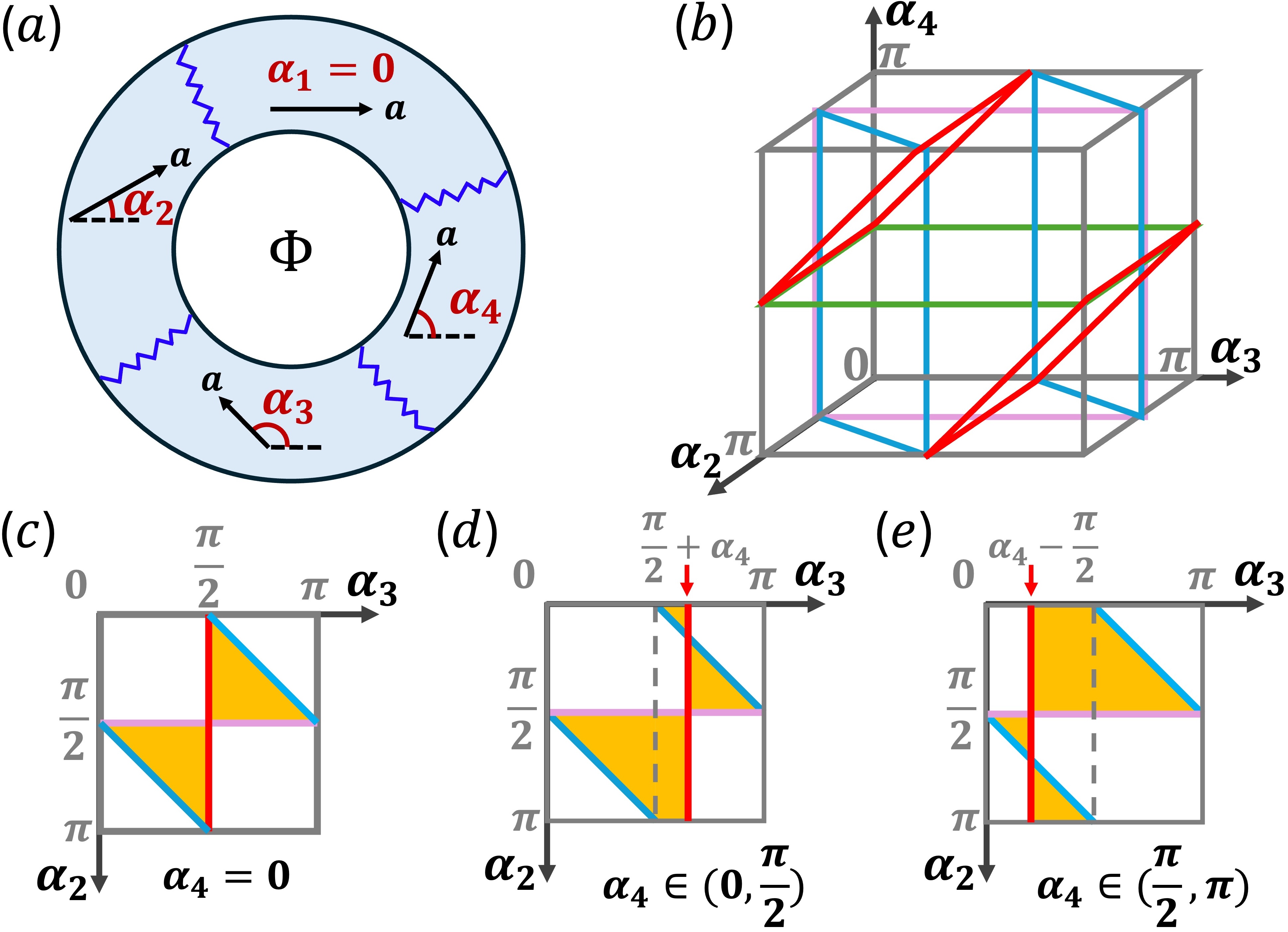}
\caption{
Four-grain superconducting ring.
(a) Geometric configuration of the ring, with grain orientations labeled by 
$\alpha_{1,2,3,4}$.
(b) Phase diagram for four-grain rings of HESP superconductor plotted in the $(\alpha_2,\alpha_3,\alpha_4)$ space with $\alpha_{1}=0$.
Colored planes mark the boundaries between $0$- and $\pi$-ring configurations.
(c-e) $2$D slices of the phase diagram at $\alpha_4=0$,
$\alpha_4\in(0,\pi/2)$ and
$\alpha_4\in(\pi/2,\pi)$, respectively.
The yellow-shaded region indicates configurations supporting spontaneous 
$\pi$ flux trapping.
The colored lines represent the intersections of these slices with the corresponding colored phase boundary planes in (b). 
}
\label{fig:4GR}
\end{figure}

Figs.~\ref{fig:3GR} and~\ref{fig:4GR} show the phase diagrams of three- and four-grain rings formed by HES superconductors.
Without loss of generality, we fix $\alpha_1=0$.
In Fig.~\ref{fig:3GR} (b), 
the red-shaded region marks the grain orientation configurations that support spontaneous 
$\pi$ flux trapping in three-grain rings.
In Fig.~\ref{fig:4GR} (c–e)
the configurations that support a $\pi$ flux trapping in four-grain rings
are highlighted in yellow 
for the representative cases of $\alpha_4 =0$, $\alpha_4 \in (0,\pi/2)$, and $\alpha_4 \in (\pi/2,\pi)$, respectively.
The phase boundaries separating $\pi$ rings from $0$ rings are shown as hyperplanes in Fig.~\ref{fig:4GR} (b).
As $N_g$ increases, the phase space becomes high dimensional and direct visualization of the phase diagram is impractical.
Assuming random and independently distributed grain orientations, we estimate the probability $P_{\pi}$ of forming a $\pi$ ring using Monte Carlo simulations.
As shown in Fig.~\ref{fig:SFig_PiRing_PiRingProb}, $P_{\pi}$ quickly approaches $1/2$ with increasing $N_g$. 
See Supplemental Material Sec.~\ref{sec:SI_PiRing} for details.

In conclusion, we analyzed Josephson couplings and spontaneous flux trapping in granular rings of unconventional superconductors, 
deriving the Josephson coupling from symmetry considerations
and systematically incorporating interface orientation disorder.
Applied to granular rings, 
we establish a no-go result that 
a single-band chiral superconductor cannot spontaneously trap magnetic flux. 
Motivated by the half-quantum flux observed in recent Little–Parks measurements on $\beta$-\ch{Bi_2Pd}, 
we examined this material in detail, incorporating its full crystalline and time-reversal symmetries. 
We propose that a fully gapped HES pairing state is consistent with all experimental results from spectroscopic and transport measurements. 
Within a minimal two-band model, 
we showed that such a state is naturally stabilized by spin–orbit coupling arising from local inversion–symmetry breaking. 
We further suggest that Little–Parks experiments on rings with controlled grain orientations,
analogous to those performed in cuprates~\cite{Tsuei1994,Tsuei2000},
would provide a direct test of this proposal.
Nevertheless, a complete understanding of how the spin–orbit coupling in $\beta$-\ch{Bi_2Pd} relates to its band topology and the emergence of topological superconductivity will require more systematic first-principles studies, which we leave for future work.

\vspace{1em}
\noindent
\emph{Acknowledgments}. 
The authors thank Yufan Li, Xiaoying Xu, Chia-Ling Chien, and Shiyuan Gao for helpful discussions. 
The authors also thank Yukio Tanaka and Ying Liu for helpful comments on the manuscript after it was posted on arXiv.
The authors acknowledge the support of NSF CAREER Grant No.~DMR-1848349. 
JYZ also acknowledges partial support from the Johns Hopkins University Theoretical Interdisciplinary Physics and Astronomy Center.

\bibliography{Ref_Manu}

\begin{thebibliography}{52}%
\makeatletter
\providecommand \@ifxundefined [1]{%
 \@ifx{#1\undefined}
}%
\providecommand \@ifnum [1]{%
 \ifnum #1\expandafter \@firstoftwo
 \else \expandafter \@secondoftwo
 \fi
}%
\providecommand \@ifx [1]{%
 \ifx #1\expandafter \@firstoftwo
 \else \expandafter \@secondoftwo
 \fi
}%
\providecommand \natexlab [1]{#1}%
\providecommand \enquote  [1]{``#1''}%
\providecommand \bibnamefont  [1]{#1}%
\providecommand \bibfnamefont [1]{#1}%
\providecommand \citenamefont [1]{#1}%
\providecommand \href@noop [0]{\@secondoftwo}%
\providecommand \href [0]{\begingroup \@sanitize@url \@href}%
\providecommand \@href[1]{\@@startlink{#1}\@@href}%
\providecommand \@@href[1]{\endgroup#1\@@endlink}%
\providecommand \@sanitize@url [0]{\catcode `\\12\catcode `\$12\catcode
  `\&12\catcode `\#12\catcode `\^12\catcode `\_12\catcode `\%12\relax}%
\providecommand \@@startlink[1]{}%
\providecommand \@@endlink[0]{}%
\providecommand \url  [0]{\begingroup\@sanitize@url \@url }%
\providecommand \@url [1]{\endgroup\@href {#1}{\urlprefix }}%
\providecommand \urlprefix  [0]{URL }%
\providecommand \Eprint [0]{\href }%
\providecommand \doibase [0]{http://dx.doi.org/}%
\providecommand \selectlanguage [0]{\@gobble}%
\providecommand \bibinfo  [0]{\@secondoftwo}%
\providecommand \bibfield  [0]{\@secondoftwo}%
\providecommand \translation [1]{[#1]}%
\providecommand \BibitemOpen [0]{}%
\providecommand \bibitemStop [0]{}%
\providecommand \bibitemNoStop [0]{.\EOS\space}%
\providecommand \EOS [0]{\spacefactor3000\relax}%
\providecommand \BibitemShut  [1]{\csname bibitem#1\endcsname}%
\let\auto@bib@innerbib\@empty
\bibitem [{\citenamefont {Josephson}(1962)}]{Josephson1962}%
  \BibitemOpen
  \bibfield  {author} {\bibinfo {author} {\bibfnamefont {B.~D.}\ \bibnamefont
  {Josephson}},\ }\bibfield  {title} {\enquote {\bibinfo {title} {Possible new
  effects in superconductive tunnelling},}\ }\href {\doibase
  https://doi.org/10.1016/0031-9163(62)91369-0} {\bibfield  {journal} {\bibinfo
   {journal} {Physics Letters}\ }\textbf {\bibinfo {volume} {1}},\ \bibinfo
  {pages} {251--253} (\bibinfo {year} {1962})}\BibitemShut {NoStop}%
\bibitem [{\citenamefont {Anderson}\ and\ \citenamefont
  {Rowell}(1963)}]{Anderson1963}%
  \BibitemOpen
  \bibfield  {author} {\bibinfo {author} {\bibfnamefont {P.~W.}\ \bibnamefont
  {Anderson}}\ and\ \bibinfo {author} {\bibfnamefont {J.~M.}\ \bibnamefont
  {Rowell}},\ }\bibfield  {title} {\enquote {\bibinfo {title} {Probable
  observation of the josephson superconducting tunneling effect},}\ }\href
  {\doibase 10.1103/PhysRevLett.10.230} {\bibfield  {journal} {\bibinfo
  {journal} {Phys. Rev. Lett.}\ }\textbf {\bibinfo {volume} {10}},\ \bibinfo
  {pages} {230--232} (\bibinfo {year} {1963})}\BibitemShut {NoStop}%
\bibitem [{\citenamefont {Jaklevic}\ \emph {et~al.}(1964)\citenamefont
  {Jaklevic}, \citenamefont {Lambe}, \citenamefont {Silver},\ and\
  \citenamefont {Mercereau}}]{Jaklevic1964}%
  \BibitemOpen
  \bibfield  {author} {\bibinfo {author} {\bibfnamefont {R.~C.}\ \bibnamefont
  {Jaklevic}}, \bibinfo {author} {\bibfnamefont {John}\ \bibnamefont {Lambe}},
  \bibinfo {author} {\bibfnamefont {A.~H.}\ \bibnamefont {Silver}}, \ and\
  \bibinfo {author} {\bibfnamefont {J.~E.}\ \bibnamefont {Mercereau}},\
  }\bibfield  {title} {\enquote {\bibinfo {title} {Quantum interference effects
  in josephson tunneling},}\ }\href {\doibase 10.1103/PhysRevLett.12.159}
  {\bibfield  {journal} {\bibinfo  {journal} {Phys. Rev. Lett.}\ }\textbf
  {\bibinfo {volume} {12}},\ \bibinfo {pages} {159--160} (\bibinfo {year}
  {1964})}\BibitemShut {NoStop}%
\bibitem [{\citenamefont {Shiba}\ and\ \citenamefont {Soda}(1969)}]{Shiba1969}%
  \BibitemOpen
  \bibfield  {author} {\bibinfo {author} {\bibfnamefont {Hiroyuki}\
  \bibnamefont {Shiba}}\ and\ \bibinfo {author} {\bibfnamefont {Toshio}\
  \bibnamefont {Soda}},\ }\bibfield  {title} {\enquote {\bibinfo {title}
  {Superconducting tunneling through the barrier with paramagnetic
  impurities},}\ }\href {\doibase 10.1143/PTP.41.25} {\bibfield  {journal}
  {\bibinfo  {journal} {Progress of Theoretical Physics}\ }\textbf {\bibinfo
  {volume} {41}},\ \bibinfo {pages} {25--44} (\bibinfo {year}
  {1969})}\BibitemShut {NoStop}%
\bibitem [{\citenamefont {Bulaevskii}\ \emph {et~al.}(1977)\citenamefont
  {Bulaevskii}, \citenamefont {Kuzii},\ and\ \citenamefont
  {Sobyanin}}]{Bulaevskii1977}%
  \BibitemOpen
  \bibfield  {author} {\bibinfo {author} {\bibfnamefont {L.~N.}\ \bibnamefont
  {Bulaevskii}}, \bibinfo {author} {\bibfnamefont {V.~V.}\ \bibnamefont
  {Kuzii}}, \ and\ \bibinfo {author} {\bibfnamefont {A.~A.}\ \bibnamefont
  {Sobyanin}},\ }\bibfield  {title} {\enquote {\bibinfo {title}
  {Superconducting systems with weak coupling to the current in the ground
  state},}\ }\href {http://jetpletters.ru/ps/1410/article_21163.pdf} {\bibfield
   {journal} {\bibinfo  {journal} {JETP Lett.}\ }\textbf {\bibinfo {volume}
  {25}},\ \bibinfo {pages} {290} (\bibinfo {year} {1977})}\BibitemShut
  {NoStop}%
\bibitem [{\citenamefont {Buzdin}\ \emph {et~al.}(1982)\citenamefont {Buzdin},
  \citenamefont {Bulaevskii},\ and\ \citenamefont {Panyukov}}]{Buzdin1982}%
  \BibitemOpen
  \bibfield  {author} {\bibinfo {author} {\bibfnamefont {A.~I.}\ \bibnamefont
  {Buzdin}}, \bibinfo {author} {\bibfnamefont {L.~N.}\ \bibnamefont
  {Bulaevskii}}, \ and\ \bibinfo {author} {\bibfnamefont {S.~V.}\ \bibnamefont
  {Panyukov}},\ }\bibfield  {title} {\enquote {\bibinfo {title}
  {Critical-current oscillations as a function of the exchange field and
  thickness of the ferromagnetic metal ({F}) in an {S-F-S} josephson
  junction},}\ }\href {http://jetpletters.ru/ps/0/article_19853.shtml}
  {\bibfield  {journal} {\bibinfo  {journal} {JETP Lett.}\ }\textbf {\bibinfo
  {volume} {35}},\ \bibinfo {pages} {178} (\bibinfo {year} {1982})}\BibitemShut
  {NoStop}%
\bibitem [{\citenamefont {Glazman}\ and\ \citenamefont
  {Matveev}(1989)}]{Glazman1989}%
  \BibitemOpen
  \bibfield  {author} {\bibinfo {author} {\bibfnamefont {L.~I.}\ \bibnamefont
  {Glazman}}\ and\ \bibinfo {author} {\bibfnamefont {K.~A.}\ \bibnamefont
  {Matveev}},\ }\bibfield  {title} {\enquote {\bibinfo {title} {Resonant
  josephson current through kondo impurities in a tunnel barrier},}\ }\href
  {http://jetpletters.ru/ps/0/article_16988.shtml} {\bibfield  {journal}
  {\bibinfo  {journal} {JETP Lett.}\ }\textbf {\bibinfo {volume} {49}},\
  \bibinfo {pages} {659} (\bibinfo {year} {1989})}\BibitemShut {NoStop}%
\bibitem [{\citenamefont {Spivak}\ and\ \citenamefont
  {Kivelson}(1991)}]{Spivak1991}%
  \BibitemOpen
  \bibfield  {author} {\bibinfo {author} {\bibfnamefont {B.~I.}\ \bibnamefont
  {Spivak}}\ and\ \bibinfo {author} {\bibfnamefont {S.~A.}\ \bibnamefont
  {Kivelson}},\ }\bibfield  {title} {\enquote {\bibinfo {title} {Negative local
  superfluid densities: The difference between dirty superconductors and dirty
  bose liquids},}\ }\href {\doibase 10.1103/PhysRevB.43.3740} {\bibfield
  {journal} {\bibinfo  {journal} {Phys. Rev. B}\ }\textbf {\bibinfo {volume}
  {43}},\ \bibinfo {pages} {3740--3743} (\bibinfo {year} {1991})}\BibitemShut
  {NoStop}%
\bibitem [{\citenamefont {Geshkenbein}\ and\ \citenamefont
  {Larkin}(1986)}]{Geshkenbein1986}%
  \BibitemOpen
  \bibfield  {author} {\bibinfo {author} {\bibfnamefont {V.~B.}\ \bibnamefont
  {Geshkenbein}}\ and\ \bibinfo {author} {\bibfnamefont {A.~I.}\ \bibnamefont
  {Larkin}},\ }\bibfield  {title} {\enquote {\bibinfo {title} {The josephson
  effect in superconductors with heavy fermions},}\ }\href@noop {} {\bibfield
  {journal} {\bibinfo  {journal} {JETP Lett.}\ }\textbf {\bibinfo {volume}
  {43}},\ \bibinfo {pages} {395} (\bibinfo {year} {1986})}\BibitemShut
  {NoStop}%
\bibitem [{\citenamefont {Geshkenbein}\ \emph {et~al.}(1987)\citenamefont
  {Geshkenbein}, \citenamefont {Larkin},\ and\ \citenamefont
  {Barone}}]{Geshkenbein1987}%
  \BibitemOpen
  \bibfield  {author} {\bibinfo {author} {\bibfnamefont {V.~B.}\ \bibnamefont
  {Geshkenbein}}, \bibinfo {author} {\bibfnamefont {A.~I.}\ \bibnamefont
  {Larkin}}, \ and\ \bibinfo {author} {\bibfnamefont {A.}~\bibnamefont
  {Barone}},\ }\bibfield  {title} {\enquote {\bibinfo {title} {Vortices with
  half magnetic flux quanta in ``heavy-fermion'' superconductors},}\ }\href
  {\doibase 10.1103/physrevb.36.235} {\bibfield  {journal} {\bibinfo  {journal}
  {Physical Review B}\ }\textbf {\bibinfo {volume} {36}},\ \bibinfo {pages}
  {235--238} (\bibinfo {year} {1987})}\BibitemShut {NoStop}%
\bibitem [{\citenamefont {Sigrist}\ and\ \citenamefont
  {M.~Rice}(1992)}]{Sigrist1992}%
  \BibitemOpen
  \bibfield  {author} {\bibinfo {author} {\bibfnamefont {Manfred}\ \bibnamefont
  {Sigrist}}\ and\ \bibinfo {author} {\bibfnamefont {T.}~\bibnamefont
  {M.~Rice}},\ }\bibfield  {title} {\enquote {\bibinfo {title} {Paramagnetic
  effect in high {$T_c$} superconductors-a hint for {$d$}-wave
  superconductivity},}\ }\href {\doibase 10.1143/JPSJ.61.4283} {\bibfield
  {journal} {\bibinfo  {journal} {Journal of the Physical Society of Japan}\
  }\textbf {\bibinfo {volume} {61}},\ \bibinfo {pages} {4283--4286} (\bibinfo
  {year} {1992})}\BibitemShut {NoStop}%
\bibitem [{\citenamefont {Sigrist}\ and\ \citenamefont
  {Rice}(1995)}]{Sigrist1995}%
  \BibitemOpen
  \bibfield  {author} {\bibinfo {author} {\bibfnamefont {Manfred}\ \bibnamefont
  {Sigrist}}\ and\ \bibinfo {author} {\bibfnamefont {T.~M.}\ \bibnamefont
  {Rice}},\ }\bibfield  {title} {\enquote {\bibinfo {title} {Unusual
  paramagnetic phenomena in granular high-temperature superconductors---a
  consequence of $d$- wave pairing?}}\ }\href {\doibase
  10.1103/RevModPhys.67.503} {\bibfield  {journal} {\bibinfo  {journal} {Rev.
  Mod. Phys.}\ }\textbf {\bibinfo {volume} {67}},\ \bibinfo {pages} {503--513}
  (\bibinfo {year} {1995})}\BibitemShut {NoStop}%
\bibitem [{\citenamefont {Wollman}\ \emph {et~al.}(1993)\citenamefont
  {Wollman}, \citenamefont {Van~Harlingen}, \citenamefont {Lee}, \citenamefont
  {Ginsberg},\ and\ \citenamefont {Leggett}}]{Wollman1993}%
  \BibitemOpen
  \bibfield  {author} {\bibinfo {author} {\bibfnamefont {D.~A.}\ \bibnamefont
  {Wollman}}, \bibinfo {author} {\bibfnamefont {D.~J.}\ \bibnamefont
  {Van~Harlingen}}, \bibinfo {author} {\bibfnamefont {W.~C.}\ \bibnamefont
  {Lee}}, \bibinfo {author} {\bibfnamefont {D.~M.}\ \bibnamefont {Ginsberg}}, \
  and\ \bibinfo {author} {\bibfnamefont {A.~J.}\ \bibnamefont {Leggett}},\
  }\bibfield  {title} {\enquote {\bibinfo {title} {Experimental determination
  of the superconducting pairing state in {YBCO} from the phase coherence of
  {YBCO}-{Pb} {dc} {SQUIDs}},}\ }\href {\doibase 10.1103/PhysRevLett.71.2134}
  {\bibfield  {journal} {\bibinfo  {journal} {Phys. Rev. Lett.}\ }\textbf
  {\bibinfo {volume} {71}},\ \bibinfo {pages} {2134--2137} (\bibinfo {year}
  {1993})}\BibitemShut {NoStop}%
\bibitem [{\citenamefont {Wollman}\ \emph {et~al.}(1995)\citenamefont
  {Wollman}, \citenamefont {Van~Harlingen}, \citenamefont {Giapintzakis},\ and\
  \citenamefont {Ginsberg}}]{Wollman1995}%
  \BibitemOpen
  \bibfield  {author} {\bibinfo {author} {\bibfnamefont {D.~A.}\ \bibnamefont
  {Wollman}}, \bibinfo {author} {\bibfnamefont {D.~J.}\ \bibnamefont
  {Van~Harlingen}}, \bibinfo {author} {\bibfnamefont {J.}~\bibnamefont
  {Giapintzakis}}, \ and\ \bibinfo {author} {\bibfnamefont {D.~M.}\
  \bibnamefont {Ginsberg}},\ }\bibfield  {title} {\enquote {\bibinfo {title}
  {Evidence for {${d}_{{x}^{2}\ensuremath{-}{y}^{2}}$} pairing from the
  magnetic field modulation of
  {Y${\mathrm{Ba}}_{2}$${\mathrm{Cu}}_{3}$${\mathrm{O}}_{7}$}-{Pb} josephson
  junctions},}\ }\href {\doibase 10.1103/PhysRevLett.74.797} {\bibfield
  {journal} {\bibinfo  {journal} {Phys. Rev. Lett.}\ }\textbf {\bibinfo
  {volume} {74}},\ \bibinfo {pages} {797--800} (\bibinfo {year}
  {1995})}\BibitemShut {NoStop}%
\bibitem [{\citenamefont {Van~Harlingen}(1995)}]{VanHarlingen1995}%
  \BibitemOpen
  \bibfield  {author} {\bibinfo {author} {\bibfnamefont {D.~J.}\ \bibnamefont
  {Van~Harlingen}},\ }\bibfield  {title} {\enquote {\bibinfo {title}
  {Phase-sensitive tests of the symmetry of the pairing state in the
  high-temperature superconductors---evidence for
  {${d}_{{x}^{2}\ensuremath{-}{y}^{2}}$} symmetry},}\ }\href {\doibase
  10.1103/RevModPhys.67.515} {\bibfield  {journal} {\bibinfo  {journal} {Rev.
  Mod. Phys.}\ }\textbf {\bibinfo {volume} {67}},\ \bibinfo {pages} {515--535}
  (\bibinfo {year} {1995})}\BibitemShut {NoStop}%
\bibitem [{\citenamefont {Strand}\ \emph {et~al.}(2009)\citenamefont {Strand},
  \citenamefont {Van~Harlingen}, \citenamefont {Kycia},\ and\ \citenamefont
  {Halperin}}]{Strand2009}%
  \BibitemOpen
  \bibfield  {author} {\bibinfo {author} {\bibfnamefont {J.~D.}\ \bibnamefont
  {Strand}}, \bibinfo {author} {\bibfnamefont {D.~J.}\ \bibnamefont
  {Van~Harlingen}}, \bibinfo {author} {\bibfnamefont {J.~B.}\ \bibnamefont
  {Kycia}}, \ and\ \bibinfo {author} {\bibfnamefont {W.~P.}\ \bibnamefont
  {Halperin}},\ }\bibfield  {title} {\enquote {\bibinfo {title} {Evidence for
  complex superconducting order parameter symmetry in the low-temperature phase
  of {${\mathrm{UPt}}_{3}$} from josephson interferometry},}\ }\href {\doibase
  10.1103/PhysRevLett.103.197002} {\bibfield  {journal} {\bibinfo  {journal}
  {Phys. Rev. Lett.}\ }\textbf {\bibinfo {volume} {103}},\ \bibinfo {pages}
  {197002} (\bibinfo {year} {2009})}\BibitemShut {NoStop}%
\bibitem [{\citenamefont {Stewart}\ \emph {et~al.}(1984)\citenamefont
  {Stewart}, \citenamefont {Fisk}, \citenamefont {Willis},\ and\ \citenamefont
  {Smith}}]{Stewart1984}%
  \BibitemOpen
  \bibfield  {author} {\bibinfo {author} {\bibfnamefont {G.~R.}\ \bibnamefont
  {Stewart}}, \bibinfo {author} {\bibfnamefont {Z.}~\bibnamefont {Fisk}},
  \bibinfo {author} {\bibfnamefont {J.~O.}\ \bibnamefont {Willis}}, \ and\
  \bibinfo {author} {\bibfnamefont {J.~L.}\ \bibnamefont {Smith}},\ }\bibfield
  {title} {\enquote {\bibinfo {title} {Possibility of coexistence of bulk
  superconductivity and spin fluctuations in {U${\mathrm{Pt}}_{3}$}},}\ }\href
  {\doibase 10.1103/PhysRevLett.52.679} {\bibfield  {journal} {\bibinfo
  {journal} {Phys. Rev. Lett.}\ }\textbf {\bibinfo {volume} {52}},\ \bibinfo
  {pages} {679--682} (\bibinfo {year} {1984})}\BibitemShut {NoStop}%
\bibitem [{\citenamefont {Aeppli}\ \emph {et~al.}(1988)\citenamefont {Aeppli},
  \citenamefont {Bucher}, \citenamefont {Broholm}, \citenamefont {Kjems},
  \citenamefont {Baumann},\ and\ \citenamefont {Hufnagl}}]{Aeppli1988}%
  \BibitemOpen
  \bibfield  {author} {\bibinfo {author} {\bibfnamefont {G.}~\bibnamefont
  {Aeppli}}, \bibinfo {author} {\bibfnamefont {E.}~\bibnamefont {Bucher}},
  \bibinfo {author} {\bibfnamefont {C.}~\bibnamefont {Broholm}}, \bibinfo
  {author} {\bibfnamefont {J.~K.}\ \bibnamefont {Kjems}}, \bibinfo {author}
  {\bibfnamefont {J.}~\bibnamefont {Baumann}}, \ and\ \bibinfo {author}
  {\bibfnamefont {J.}~\bibnamefont {Hufnagl}},\ }\bibfield  {title} {\enquote
  {\bibinfo {title} {Magnetic order and fluctuations in superconducting
  {${\mathrm{UPt}}_{3}$}},}\ }\href {\doibase 10.1103/PhysRevLett.60.615}
  {\bibfield  {journal} {\bibinfo  {journal} {Phys. Rev. Lett.}\ }\textbf
  {\bibinfo {volume} {60}},\ \bibinfo {pages} {615--618} (\bibinfo {year}
  {1988})}\BibitemShut {NoStop}%
\bibitem [{\citenamefont {Aeppli}\ \emph {et~al.}(1989)\citenamefont {Aeppli},
  \citenamefont {Bishop}, \citenamefont {Broholm}, \citenamefont {Bucher},
  \citenamefont {Siemensmeyer}, \citenamefont {Steiner},\ and\ \citenamefont
  {St\"usser}}]{Aeppli1989}%
  \BibitemOpen
  \bibfield  {author} {\bibinfo {author} {\bibfnamefont {G.}~\bibnamefont
  {Aeppli}}, \bibinfo {author} {\bibfnamefont {D.}~\bibnamefont {Bishop}},
  \bibinfo {author} {\bibfnamefont {C.}~\bibnamefont {Broholm}}, \bibinfo
  {author} {\bibfnamefont {E.}~\bibnamefont {Bucher}}, \bibinfo {author}
  {\bibfnamefont {K.}~\bibnamefont {Siemensmeyer}}, \bibinfo {author}
  {\bibfnamefont {M.}~\bibnamefont {Steiner}}, \ and\ \bibinfo {author}
  {\bibfnamefont {N.}~\bibnamefont {St\"usser}},\ }\bibfield  {title} {\enquote
  {\bibinfo {title} {Magnetic order in the different superconducting states of
  {${\mathrm{UPt}}_{3}$}},}\ }\href {\doibase 10.1103/PhysRevLett.63.676}
  {\bibfield  {journal} {\bibinfo  {journal} {Phys. Rev. Lett.}\ }\textbf
  {\bibinfo {volume} {63}},\ \bibinfo {pages} {676--679} (\bibinfo {year}
  {1989})}\BibitemShut {NoStop}%
\bibitem [{\citenamefont {Fisher}\ \emph {et~al.}(1989)\citenamefont {Fisher},
  \citenamefont {Kim}, \citenamefont {Woodfield}, \citenamefont {Phillips},
  \citenamefont {Taillefer}, \citenamefont {Hasselbach}, \citenamefont
  {Flouquet}, \citenamefont {Giorgi},\ and\ \citenamefont
  {Smith}}]{Fisher1989}%
  \BibitemOpen
  \bibfield  {author} {\bibinfo {author} {\bibfnamefont {R.~A.}\ \bibnamefont
  {Fisher}}, \bibinfo {author} {\bibfnamefont {S.}~\bibnamefont {Kim}},
  \bibinfo {author} {\bibfnamefont {B.~F.}\ \bibnamefont {Woodfield}}, \bibinfo
  {author} {\bibfnamefont {N.~E.}\ \bibnamefont {Phillips}}, \bibinfo {author}
  {\bibfnamefont {L.}~\bibnamefont {Taillefer}}, \bibinfo {author}
  {\bibfnamefont {K.}~\bibnamefont {Hasselbach}}, \bibinfo {author}
  {\bibfnamefont {J.}~\bibnamefont {Flouquet}}, \bibinfo {author}
  {\bibfnamefont {A.~L.}\ \bibnamefont {Giorgi}}, \ and\ \bibinfo {author}
  {\bibfnamefont {J.~L.}\ \bibnamefont {Smith}},\ }\bibfield  {title} {\enquote
  {\bibinfo {title} {Specific heat of {${\mathrm{UPt}}_{3}$}: Evidence for
  unconventional superconductivity},}\ }\href {\doibase
  10.1103/PhysRevLett.62.1411} {\bibfield  {journal} {\bibinfo  {journal}
  {Phys. Rev. Lett.}\ }\textbf {\bibinfo {volume} {62}},\ \bibinfo {pages}
  {1411--1414} (\bibinfo {year} {1989})}\BibitemShut {NoStop}%
\bibitem [{\citenamefont {Sauls}(1994)}]{Sauls1994}%
  \BibitemOpen
  \bibfield  {author} {\bibinfo {author} {\bibfnamefont {J.A.}\ \bibnamefont
  {Sauls}},\ }\bibfield  {title} {\enquote {\bibinfo {title} {The order
  parameter for the superconducting phases of {$\mathrm{UPt}_3$}},}\ }\href
  {\doibase 10.1080/00018739400101475} {\bibfield  {journal} {\bibinfo
  {journal} {Advances in Physics}\ }\textbf {\bibinfo {volume} {43}},\ \bibinfo
  {pages} {113--141} (\bibinfo {year} {1994})}\BibitemShut {NoStop}%
\bibitem [{\citenamefont {Tou}\ \emph {et~al.}(1996)\citenamefont {Tou},
  \citenamefont {Kitaoka}, \citenamefont {Asayama}, \citenamefont {Kimura},
  \citenamefont {\ifmmode~\bar{O}\else \={O}\fi{}nuki}, \citenamefont
  {Yamamoto},\ and\ \citenamefont {Maezawa}}]{Tou1996}%
  \BibitemOpen
  \bibfield  {author} {\bibinfo {author} {\bibfnamefont {H.}~\bibnamefont
  {Tou}}, \bibinfo {author} {\bibfnamefont {Y.}~\bibnamefont {Kitaoka}},
  \bibinfo {author} {\bibfnamefont {K.}~\bibnamefont {Asayama}}, \bibinfo
  {author} {\bibfnamefont {N.}~\bibnamefont {Kimura}}, \bibinfo {author}
  {\bibfnamefont {Y.}~\bibnamefont {\ifmmode~\bar{O}\else \={O}\fi{}nuki}},
  \bibinfo {author} {\bibfnamefont {E.}~\bibnamefont {Yamamoto}}, \ and\
  \bibinfo {author} {\bibfnamefont {K.}~\bibnamefont {Maezawa}},\ }\bibfield
  {title} {\enquote {\bibinfo {title} {Odd-parity superconductivity with
  parallel spin pairing in {${\mathrm{UPt}}_{3}$}: Evidence from
  {${}^{195}\mathrm{Pt}$} knight shift study},}\ }\href {\doibase
  10.1103/PhysRevLett.77.1374} {\bibfield  {journal} {\bibinfo  {journal}
  {Phys. Rev. Lett.}\ }\textbf {\bibinfo {volume} {77}},\ \bibinfo {pages}
  {1374--1377} (\bibinfo {year} {1996})}\BibitemShut {NoStop}%
\bibitem [{\citenamefont {Park}\ and\ \citenamefont {Joynt}(1996)}]{Park1996}%
  \BibitemOpen
  \bibfield  {author} {\bibinfo {author} {\bibfnamefont {K.~A.}\ \bibnamefont
  {Park}}\ and\ \bibinfo {author} {\bibfnamefont {Robert}\ \bibnamefont
  {Joynt}},\ }\bibfield  {title} {\enquote {\bibinfo {title}
  {{${\mathit{E}}_{1\mathit{g}}$} model of superconducting
  {${\mathrm{UPt}}_{3}$}},}\ }\href {\doibase 10.1103/PhysRevB.53.12346}
  {\bibfield  {journal} {\bibinfo  {journal} {Phys. Rev. B}\ }\textbf {\bibinfo
  {volume} {53}},\ \bibinfo {pages} {12346--12363} (\bibinfo {year}
  {1996})}\BibitemShut {NoStop}%
\bibitem [{\citenamefont {Joynt}\ and\ \citenamefont
  {Taillefer}(2002)}]{Joynt2002}%
  \BibitemOpen
  \bibfield  {author} {\bibinfo {author} {\bibfnamefont {Robert}\ \bibnamefont
  {Joynt}}\ and\ \bibinfo {author} {\bibfnamefont {Louis}\ \bibnamefont
  {Taillefer}},\ }\bibfield  {title} {\enquote {\bibinfo {title} {The
  superconducting phases of {${\mathrm{UPt}}_{3}$}},}\ }\href {\doibase
  10.1103/RevModPhys.74.235} {\bibfield  {journal} {\bibinfo  {journal} {Rev.
  Mod. Phys.}\ }\textbf {\bibinfo {volume} {74}},\ \bibinfo {pages} {235--294}
  (\bibinfo {year} {2002})}\BibitemShut {NoStop}%
\bibitem [{\citenamefont {Tsuei}\ \emph {et~al.}(1994)\citenamefont {Tsuei},
  \citenamefont {Kirtley}, \citenamefont {Chi}, \citenamefont {Yu-Jahnes},
  \citenamefont {Gupta}, \citenamefont {Shaw}, \citenamefont {Sun},\ and\
  \citenamefont {Ketchen}}]{Tsuei1994}%
  \BibitemOpen
  \bibfield  {author} {\bibinfo {author} {\bibfnamefont {C.~C.}\ \bibnamefont
  {Tsuei}}, \bibinfo {author} {\bibfnamefont {J.~R.}\ \bibnamefont {Kirtley}},
  \bibinfo {author} {\bibfnamefont {C.~C.}\ \bibnamefont {Chi}}, \bibinfo
  {author} {\bibfnamefont {Lock~See}\ \bibnamefont {Yu-Jahnes}}, \bibinfo
  {author} {\bibfnamefont {A.}~\bibnamefont {Gupta}}, \bibinfo {author}
  {\bibfnamefont {T.}~\bibnamefont {Shaw}}, \bibinfo {author} {\bibfnamefont
  {J.~Z.}\ \bibnamefont {Sun}}, \ and\ \bibinfo {author} {\bibfnamefont
  {M.~B.}\ \bibnamefont {Ketchen}},\ }\bibfield  {title} {\enquote {\bibinfo
  {title} {Pairing symmetry and flux quantization in a tricrystal
  superconducting ring of
  {$\mathrm{Y}{\mathrm{Ba}}_{2}{\mathrm{Cu}}_{3}{\mathrm{O}}_{7\ensuremath{-}\ensuremath{\delta}}$}},}\
  }\href {\doibase 10.1103/PhysRevLett.73.593} {\bibfield  {journal} {\bibinfo
  {journal} {Phys. Rev. Lett.}\ }\textbf {\bibinfo {volume} {73}},\ \bibinfo
  {pages} {593--596} (\bibinfo {year} {1994})}\BibitemShut {NoStop}%
\bibitem [{\citenamefont {Kirtley}\ \emph {et~al.}(1996)\citenamefont
  {Kirtley}, \citenamefont {Tsuei}, \citenamefont {Raffy}, \citenamefont {Li},
  \citenamefont {Gupta}, \citenamefont {Sun},\ and\ \citenamefont
  {Megtert}}]{Kirtley1996}%
  \BibitemOpen
  \bibfield  {author} {\bibinfo {author} {\bibfnamefont {J.~R.}\ \bibnamefont
  {Kirtley}}, \bibinfo {author} {\bibfnamefont {C.~C.}\ \bibnamefont {Tsuei}},
  \bibinfo {author} {\bibfnamefont {H.}~\bibnamefont {Raffy}}, \bibinfo
  {author} {\bibfnamefont {Z.~Z.}\ \bibnamefont {Li}}, \bibinfo {author}
  {\bibfnamefont {A.}~\bibnamefont {Gupta}}, \bibinfo {author} {\bibfnamefont
  {J.~Z.}\ \bibnamefont {Sun}}, \ and\ \bibinfo {author} {\bibfnamefont
  {S.}~\bibnamefont {Megtert}},\ }\bibfield  {title} {\enquote {\bibinfo
  {title} {Half-integer flux quantum effect in tricrystal
  {Bi$_2$Sr$_2$CaCu$_2$O$_{8+\delta}$}},}\ }\href {\doibase
  10.1209/epl/i1996-00291-y} {\bibfield  {journal} {\bibinfo  {journal}
  {Europhysics Letters}\ }\textbf {\bibinfo {volume} {36}},\ \bibinfo {pages}
  {707} (\bibinfo {year} {1996})}\BibitemShut {NoStop}%
\bibitem [{\citenamefont {Tsuei}\ and\ \citenamefont
  {Kirtley}(2000)}]{Tsuei2000}%
  \BibitemOpen
  \bibfield  {author} {\bibinfo {author} {\bibfnamefont {C.~C.}\ \bibnamefont
  {Tsuei}}\ and\ \bibinfo {author} {\bibfnamefont {J.~R.}\ \bibnamefont
  {Kirtley}},\ }\bibfield  {title} {\enquote {\bibinfo {title} {Pairing
  symmetry in cuprate superconductors},}\ }\href {\doibase
  10.1103/RevModPhys.72.969} {\bibfield  {journal} {\bibinfo  {journal} {Rev.
  Mod. Phys.}\ }\textbf {\bibinfo {volume} {72}},\ \bibinfo {pages} {969--1016}
  (\bibinfo {year} {2000})}\BibitemShut {NoStop}%
\bibitem [{\citenamefont {Little}\ and\ \citenamefont
  {Parks}(1962)}]{Little1962}%
  \BibitemOpen
  \bibfield  {author} {\bibinfo {author} {\bibfnamefont {W.~A.}\ \bibnamefont
  {Little}}\ and\ \bibinfo {author} {\bibfnamefont {R.~D.}\ \bibnamefont
  {Parks}},\ }\bibfield  {title} {\enquote {\bibinfo {title} {Observation of
  quantum periodicity in the transition temperature of a superconducting
  cylinder},}\ }\href {\doibase 10.1103/PhysRevLett.9.9} {\bibfield  {journal}
  {\bibinfo  {journal} {Phys. Rev. Lett.}\ }\textbf {\bibinfo {volume} {9}},\
  \bibinfo {pages} {9--12} (\bibinfo {year} {1962})}\BibitemShut {NoStop}%
\bibitem [{\citenamefont {Ge}\ \emph {et~al.}(2024)\citenamefont {Ge},
  \citenamefont {Wang}, \citenamefont {Xing}, \citenamefont {Yin},
  \citenamefont {Wang}, \citenamefont {Shen}, \citenamefont {Lei},
  \citenamefont {Wang},\ and\ \citenamefont {Wang}}]{Ge2024}%
  \BibitemOpen
  \bibfield  {author} {\bibinfo {author} {\bibfnamefont {Jun}\ \bibnamefont
  {Ge}}, \bibinfo {author} {\bibfnamefont {Pinyuan}\ \bibnamefont {Wang}},
  \bibinfo {author} {\bibfnamefont {Ying}\ \bibnamefont {Xing}}, \bibinfo
  {author} {\bibfnamefont {Qiangwei}\ \bibnamefont {Yin}}, \bibinfo {author}
  {\bibfnamefont {Anqi}\ \bibnamefont {Wang}}, \bibinfo {author} {\bibfnamefont
  {Jie}\ \bibnamefont {Shen}}, \bibinfo {author} {\bibfnamefont {Hechang}\
  \bibnamefont {Lei}}, \bibinfo {author} {\bibfnamefont {Ziqiang}\ \bibnamefont
  {Wang}}, \ and\ \bibinfo {author} {\bibfnamefont {Jian}\ \bibnamefont
  {Wang}},\ }\bibfield  {title} {\enquote {\bibinfo {title} {Charge-$4e$ and
  charge-$6e$ flux quantization and higher charge superconductivity in kagome
  superconductor ring devices},}\ }\href {\doibase 10.1103/PhysRevX.14.021025}
  {\bibfield  {journal} {\bibinfo  {journal} {Phys. Rev. X}\ }\textbf {\bibinfo
  {volume} {14}},\ \bibinfo {pages} {021025} (\bibinfo {year}
  {2024})}\BibitemShut {NoStop}%
\bibitem [{\citenamefont {Zhang}\ \emph {et~al.}(2024)\citenamefont {Zhang},
  \citenamefont {Wang},\ and\ \citenamefont {Hu}}]{Zhang2024}%
  \BibitemOpen
  \bibfield  {author} {\bibinfo {author} {\bibfnamefont {Ling-Feng}\
  \bibnamefont {Zhang}}, \bibinfo {author} {\bibfnamefont {Zhi}\ \bibnamefont
  {Wang}}, \ and\ \bibinfo {author} {\bibfnamefont {Xiao}\ \bibnamefont {Hu}},\
  }\bibfield  {title} {\enquote {\bibinfo {title} {{Higgs}-{Leggett} mechanism
  for the elusive {$\phi_0/3=hc/6e$} oscillation in {Little}-{Parks} setup of
  kagome superconductor {$\ch{CsV_3Sb_5}$}},}\ }\href {\doibase
  10.1038/s42005-024-01663-0} {\bibfield  {journal} {\bibinfo  {journal}
  {Communications Physics}\ }\textbf {\bibinfo {volume} {7}},\ \bibinfo {pages}
  {210} (\bibinfo {year} {2024})}\BibitemShut {NoStop}%
\bibitem [{\citenamefont {Pan}\ \emph {et~al.}(2024)\citenamefont {Pan},
  \citenamefont {Lu}, \citenamefont {Yang},\ and\ \citenamefont
  {Wu}}]{Pan2024}%
  \BibitemOpen
  \bibfield  {author} {\bibinfo {author} {\bibfnamefont {Zhiming}\ \bibnamefont
  {Pan}}, \bibinfo {author} {\bibfnamefont {Chen}\ \bibnamefont {Lu}}, \bibinfo
  {author} {\bibfnamefont {Fan}\ \bibnamefont {Yang}}, \ and\ \bibinfo {author}
  {\bibfnamefont {Congjun}\ \bibnamefont {Wu}},\ }\bibfield  {title} {\enquote
  {\bibinfo {title} {Frustrated superconductivity and sextetting order},}\
  }\href {\doibase 10.1007/s11433-024-2396-y} {\bibfield  {journal} {\bibinfo
  {journal} {Science China Physics, Mechanics \& Astronomy}\ }\textbf {\bibinfo
  {volume} {67}},\ \bibinfo {pages} {287412} (\bibinfo {year}
  {2024})}\BibitemShut {NoStop}%
\bibitem [{\citenamefont {Almoalem}\ \emph {et~al.}(2024)\citenamefont
  {Almoalem}, \citenamefont {Feldman}, \citenamefont {Mangel}, \citenamefont
  {Shlafman}, \citenamefont {Yaish}, \citenamefont {Fischer}, \citenamefont
  {Moshe}, \citenamefont {Ruhman},\ and\ \citenamefont
  {Kanigel}}]{Almoalem2024}%
  \BibitemOpen
  \bibfield  {author} {\bibinfo {author} {\bibfnamefont {Avior}\ \bibnamefont
  {Almoalem}}, \bibinfo {author} {\bibfnamefont {Irena}\ \bibnamefont
  {Feldman}}, \bibinfo {author} {\bibfnamefont {Ilay}\ \bibnamefont {Mangel}},
  \bibinfo {author} {\bibfnamefont {Michael}\ \bibnamefont {Shlafman}},
  \bibinfo {author} {\bibfnamefont {Yuval~E.}\ \bibnamefont {Yaish}}, \bibinfo
  {author} {\bibfnamefont {Mark~H.}\ \bibnamefont {Fischer}}, \bibinfo {author}
  {\bibfnamefont {Michael}\ \bibnamefont {Moshe}}, \bibinfo {author}
  {\bibfnamefont {Jonathan}\ \bibnamefont {Ruhman}}, \ and\ \bibinfo {author}
  {\bibfnamefont {Amit}\ \bibnamefont {Kanigel}},\ }\bibfield  {title}
  {\enquote {\bibinfo {title} {The observation of $\pi$-shifts in the
  little-parks effect in {4Hb-TaS$_2$}},}\ }\href {\doibase
  10.1038/s41467-024-48260-x} {\bibfield  {journal} {\bibinfo  {journal}
  {Nature Communications}\ }\textbf {\bibinfo {volume} {15}},\ \bibinfo {pages}
  {4623} (\bibinfo {year} {2024})}\BibitemShut {NoStop}%
\bibitem [{\citenamefont {Fischer}\ \emph
  {et~al.}(2023{\natexlab{a}})\citenamefont {Fischer}, \citenamefont {Lee},\
  and\ \citenamefont {Ruhman}}]{Fischer2023b}%
  \BibitemOpen
  \bibfield  {author} {\bibinfo {author} {\bibfnamefont {Mark~H.}\ \bibnamefont
  {Fischer}}, \bibinfo {author} {\bibfnamefont {Patrick~A.}\ \bibnamefont
  {Lee}}, \ and\ \bibinfo {author} {\bibfnamefont {Jonathan}\ \bibnamefont
  {Ruhman}},\ }\bibfield  {title} {\enquote {\bibinfo {title} {Mechanism for
  $\ensuremath{\pi}$ phase shifts in {Little}-{Parks} experiments: Application
  to {$4Hb\text{\ensuremath{-}}{\mathrm{TaS}}_{2}$} and to
  {$2H\text{\ensuremath{-}}{\mathrm{TaS}}_{2}$} intercalated with chiral
  molecules},}\ }\href {\doibase 10.1103/PhysRevB.108.L180505} {\bibfield
  {journal} {\bibinfo  {journal} {Phys. Rev. B}\ }\textbf {\bibinfo {volume}
  {108}},\ \bibinfo {pages} {L180505} (\bibinfo {year}
  {2023}{\natexlab{a}})}\BibitemShut {NoStop}%
\bibitem [{\citenamefont {Li}\ \emph {et~al.}(2019)\citenamefont {Li},
  \citenamefont {Xu}, \citenamefont {Lee}, \citenamefont {Chu},\ and\
  \citenamefont {Chien}}]{Li2019}%
  \BibitemOpen
  \bibfield  {author} {\bibinfo {author} {\bibfnamefont {Yufan}\ \bibnamefont
  {Li}}, \bibinfo {author} {\bibfnamefont {Xiaoying}\ \bibnamefont {Xu}},
  \bibinfo {author} {\bibfnamefont {M.-H.}\ \bibnamefont {Lee}}, \bibinfo
  {author} {\bibfnamefont {M.-W.}\ \bibnamefont {Chu}}, \ and\ \bibinfo
  {author} {\bibfnamefont {C.~L.}\ \bibnamefont {Chien}},\ }\bibfield  {title}
  {\enquote {\bibinfo {title} {Observation of half-quantum flux in the
  unconventional superconductor {$\beta$-$\ch{Bi_2Pd}$}},}\ }\href {\doibase
  10.1126/science.aau6539} {\bibfield  {journal} {\bibinfo  {journal}
  {Science}\ }\textbf {\bibinfo {volume} {366}},\ \bibinfo {pages} {238--241}
  (\bibinfo {year} {2019})}\BibitemShut {NoStop}%
\bibitem [{\citenamefont {Xu}\ \emph {et~al.}(2024)\citenamefont {Xu},
  \citenamefont {Li},\ and\ \citenamefont {Chien}}]{Xu2024}%
  \BibitemOpen
  \bibfield  {author} {\bibinfo {author} {\bibfnamefont {Xiaoying}\
  \bibnamefont {Xu}}, \bibinfo {author} {\bibfnamefont {Yufan}\ \bibnamefont
  {Li}}, \ and\ \bibinfo {author} {\bibfnamefont {C.~L.}\ \bibnamefont
  {Chien}},\ }\bibfield  {title} {\enquote {\bibinfo {title} {Observation of
  odd-parity superconductivity with the {Geshkenbein-Larkin-Barone} composite
  rings},}\ }\href {\doibase 10.1103/PhysRevLett.132.056001} {\bibfield
  {journal} {\bibinfo  {journal} {Phys. Rev. Lett.}\ }\textbf {\bibinfo
  {volume} {132}},\ \bibinfo {pages} {056001} (\bibinfo {year}
  {2024})}\BibitemShut {NoStop}%
\bibitem [{\citenamefont {Li}\ \emph {et~al.}(2024)\citenamefont {Li},
  \citenamefont {Xu}, \citenamefont {Lee},\ and\ \citenamefont
  {Chien}}]{Li2024}%
  \BibitemOpen
  \bibfield  {author} {\bibinfo {author} {\bibfnamefont {Yufan}\ \bibnamefont
  {Li}}, \bibinfo {author} {\bibfnamefont {Xiaoying}\ \bibnamefont {Xu}},
  \bibinfo {author} {\bibfnamefont {Shu-Ping}\ \bibnamefont {Lee}}, \ and\
  \bibinfo {author} {\bibfnamefont {C.~L.}\ \bibnamefont {Chien}},\ }\bibfield
  {title} {\enquote {\bibinfo {title} {Unconventional periodicities of the
  {Little-Parks} effect observed in a topological superconductor},}\ }\href
  {\doibase 10.1103/PhysRevB.109.L060504} {\bibfield  {journal} {\bibinfo
  {journal} {Phys. Rev. B}\ }\textbf {\bibinfo {volume} {109}},\ \bibinfo
  {pages} {L060504} (\bibinfo {year} {2024})}\BibitemShut {NoStop}%
\bibitem [{\citenamefont {Herrera}\ \emph {et~al.}(2015)\citenamefont
  {Herrera}, \citenamefont {Guillam\'on}, \citenamefont {Galvis}, \citenamefont
  {Correa}, \citenamefont {Fente}, \citenamefont {Luccas}, \citenamefont
  {Mompean}, \citenamefont {Garc\'{\i}a-Hern\'andez}, \citenamefont {Vieira},
  \citenamefont {Brison},\ and\ \citenamefont {Suderow}}]{Herrera2015}%
  \BibitemOpen
  \bibfield  {author} {\bibinfo {author} {\bibfnamefont {E.}~\bibnamefont
  {Herrera}}, \bibinfo {author} {\bibfnamefont {I.}~\bibnamefont
  {Guillam\'on}}, \bibinfo {author} {\bibfnamefont {J.~A.}\ \bibnamefont
  {Galvis}}, \bibinfo {author} {\bibfnamefont {A.}~\bibnamefont {Correa}},
  \bibinfo {author} {\bibfnamefont {A.}~\bibnamefont {Fente}}, \bibinfo
  {author} {\bibfnamefont {R.~F.}\ \bibnamefont {Luccas}}, \bibinfo {author}
  {\bibfnamefont {F.~J.}\ \bibnamefont {Mompean}}, \bibinfo {author}
  {\bibfnamefont {M.}~\bibnamefont {Garc\'{\i}a-Hern\'andez}}, \bibinfo
  {author} {\bibfnamefont {S.}~\bibnamefont {Vieira}}, \bibinfo {author}
  {\bibfnamefont {J.~P.}\ \bibnamefont {Brison}}, \ and\ \bibinfo {author}
  {\bibfnamefont {H.}~\bibnamefont {Suderow}},\ }\bibfield  {title} {\enquote
  {\bibinfo {title} {Magnetic field dependence of the density of states in the
  multiband superconductor {$\ensuremath{\beta}$-Bi$_{2}$Pd}},}\ }\href
  {\doibase 10.1103/PhysRevB.92.054507} {\bibfield  {journal} {\bibinfo
  {journal} {Phys. Rev. B}\ }\textbf {\bibinfo {volume} {92}},\ \bibinfo
  {pages} {054507} (\bibinfo {year} {2015})}\BibitemShut {NoStop}%
\bibitem [{\citenamefont {Lv}\ \emph {et~al.}(2017)\citenamefont {Lv},
  \citenamefont {Wang}, \citenamefont {Zhang}, \citenamefont {Ding},
  \citenamefont {Li}, \citenamefont {Wang}, \citenamefont {He}, \citenamefont
  {Song}, \citenamefont {Ma},\ and\ \citenamefont {Xue}}]{Lv2017}%
  \BibitemOpen
  \bibfield  {author} {\bibinfo {author} {\bibfnamefont {Yan-Feng}\
  \bibnamefont {Lv}}, \bibinfo {author} {\bibfnamefont {Wen-Lin}\ \bibnamefont
  {Wang}}, \bibinfo {author} {\bibfnamefont {Yi-Min}\ \bibnamefont {Zhang}},
  \bibinfo {author} {\bibfnamefont {Hao}\ \bibnamefont {Ding}}, \bibinfo
  {author} {\bibfnamefont {Wei}\ \bibnamefont {Li}}, \bibinfo {author}
  {\bibfnamefont {Lili}\ \bibnamefont {Wang}}, \bibinfo {author} {\bibfnamefont
  {Ke}~\bibnamefont {He}}, \bibinfo {author} {\bibfnamefont {Can-Li}\
  \bibnamefont {Song}}, \bibinfo {author} {\bibfnamefont {Xu-Cun}\ \bibnamefont
  {Ma}}, \ and\ \bibinfo {author} {\bibfnamefont {Qi-Kun}\ \bibnamefont
  {Xue}},\ }\bibfield  {title} {\enquote {\bibinfo {title} {Experimental
  signature of topological superconductivity and {Majorana} zero modes on
  {$\beta$-$\mathrm{Bi}_2\mathrm{Pd}$} thin films},}\ }\href {\doibase
  https://doi.org/10.1016/j.scib.2017.05.008} {\bibfield  {journal} {\bibinfo
  {journal} {Science Bulletin}\ }\textbf {\bibinfo {volume} {62}},\ \bibinfo
  {pages} {852--856} (\bibinfo {year} {2017})}\BibitemShut {NoStop}%
\bibitem [{\citenamefont {Ka\ifmmode \check{c}\else
  \v{c}\fi{}mar\ifmmode~\check{c}\else \v{c}\fi{}\'{\i}k}\ \emph
  {et~al.}(2016)\citenamefont {Ka\ifmmode \check{c}\else
  \v{c}\fi{}mar\ifmmode~\check{c}\else \v{c}\fi{}\'{\i}k}, \citenamefont
  {Pribulov\'a}, \citenamefont {Samuely}, \citenamefont {Szab\'o},
  \citenamefont {Cambel}, \citenamefont {\ifmmode~\check{S}\else
  \v{S}\fi{}olt\'ys}, \citenamefont {Herrera}, \citenamefont {Suderow},
  \citenamefont {Correa-Orellana}, \citenamefont {Prabhakaran},\ and\
  \citenamefont {Samuely}}]{Kacmarcik2016}%
  \BibitemOpen
  \bibfield  {author} {\bibinfo {author} {\bibfnamefont {J.}~\bibnamefont
  {Ka\ifmmode \check{c}\else \v{c}\fi{}mar\ifmmode~\check{c}\else
  \v{c}\fi{}\'{\i}k}}, \bibinfo {author} {\bibfnamefont {Z.}~\bibnamefont
  {Pribulov\'a}}, \bibinfo {author} {\bibfnamefont {T.}~\bibnamefont
  {Samuely}}, \bibinfo {author} {\bibfnamefont {P.}~\bibnamefont {Szab\'o}},
  \bibinfo {author} {\bibfnamefont {V.}~\bibnamefont {Cambel}}, \bibinfo
  {author} {\bibfnamefont {J.}~\bibnamefont {\ifmmode~\check{S}\else
  \v{S}\fi{}olt\'ys}}, \bibinfo {author} {\bibfnamefont {E.}~\bibnamefont
  {Herrera}}, \bibinfo {author} {\bibfnamefont {H.}~\bibnamefont {Suderow}},
  \bibinfo {author} {\bibfnamefont {A.}~\bibnamefont {Correa-Orellana}},
  \bibinfo {author} {\bibfnamefont {D.}~\bibnamefont {Prabhakaran}}, \ and\
  \bibinfo {author} {\bibfnamefont {P.}~\bibnamefont {Samuely}},\ }\bibfield
  {title} {\enquote {\bibinfo {title} {Single-gap superconductivity in
  $\ensuremath{\beta}\text{\ensuremath{-}}\mathrm{B}{\mathrm{i}}_{2}\mathrm{Pd}$},}\
  }\href {\doibase 10.1103/PhysRevB.93.144502} {\bibfield  {journal} {\bibinfo
  {journal} {Phys. Rev. B}\ }\textbf {\bibinfo {volume} {93}},\ \bibinfo
  {pages} {144502} (\bibinfo {year} {2016})}\BibitemShut {NoStop}%
\bibitem [{\citenamefont {Chen}\ \emph {et~al.}(2020)\citenamefont {Chen},
  \citenamefont {Wang}, \citenamefont {Pang}, \citenamefont {Su}, \citenamefont
  {Chen},\ and\ \citenamefont {Yuan}}]{Chen2020}%
  \BibitemOpen
  \bibfield  {author} {\bibinfo {author} {\bibfnamefont {Jian}\ \bibnamefont
  {Chen}}, \bibinfo {author} {\bibfnamefont {An}~\bibnamefont {Wang}}, \bibinfo
  {author} {\bibfnamefont {Guiming}\ \bibnamefont {Pang}}, \bibinfo {author}
  {\bibfnamefont {Hang}\ \bibnamefont {Su}}, \bibinfo {author} {\bibfnamefont
  {Ye}~\bibnamefont {Chen}}, \ and\ \bibinfo {author} {\bibfnamefont {Huiqiu}\
  \bibnamefont {Yuan}},\ }\bibfield  {title} {\enquote {\bibinfo {title}
  {Nodeless superconductivity in
  {$\ensuremath{\beta}\text{\ensuremath{-}}{\mathrm{PdBi}}_{2}$}},}\ }\href
  {\doibase 10.1103/PhysRevB.101.054514} {\bibfield  {journal} {\bibinfo
  {journal} {Phys. Rev. B}\ }\textbf {\bibinfo {volume} {101}},\ \bibinfo
  {pages} {054514} (\bibinfo {year} {2020})}\BibitemShut {NoStop}%
\bibitem [{\citenamefont {Biswas}\ \emph {et~al.}(2016)\citenamefont {Biswas},
  \citenamefont {Mazzone}, \citenamefont {Sibille}, \citenamefont
  {Pomjakushina}, \citenamefont {Conder}, \citenamefont {Luetkens},
  \citenamefont {Baines}, \citenamefont {Gavilano}, \citenamefont {Kenzelmann},
  \citenamefont {Amato},\ and\ \citenamefont {Morenzoni}}]{Biswas2016}%
  \BibitemOpen
  \bibfield  {author} {\bibinfo {author} {\bibfnamefont {P.~K.}\ \bibnamefont
  {Biswas}}, \bibinfo {author} {\bibfnamefont {D.~G.}\ \bibnamefont {Mazzone}},
  \bibinfo {author} {\bibfnamefont {R.}~\bibnamefont {Sibille}}, \bibinfo
  {author} {\bibfnamefont {E.}~\bibnamefont {Pomjakushina}}, \bibinfo {author}
  {\bibfnamefont {K.}~\bibnamefont {Conder}}, \bibinfo {author} {\bibfnamefont
  {H.}~\bibnamefont {Luetkens}}, \bibinfo {author} {\bibfnamefont
  {C.}~\bibnamefont {Baines}}, \bibinfo {author} {\bibfnamefont {J.~L.}\
  \bibnamefont {Gavilano}}, \bibinfo {author} {\bibfnamefont {M.}~\bibnamefont
  {Kenzelmann}}, \bibinfo {author} {\bibfnamefont {A.}~\bibnamefont {Amato}}, \
  and\ \bibinfo {author} {\bibfnamefont {E.}~\bibnamefont {Morenzoni}},\
  }\bibfield  {title} {\enquote {\bibinfo {title} {Fully gapped
  superconductivity in the topological superconductor
  {$\ensuremath{\beta}\ensuremath{-}{\mathrm{PdBi}}_{2}$}},}\ }\href {\doibase
  10.1103/PhysRevB.93.220504} {\bibfield  {journal} {\bibinfo  {journal} {Phys.
  Rev. B}\ }\textbf {\bibinfo {volume} {93}},\ \bibinfo {pages} {220504}
  (\bibinfo {year} {2016})}\BibitemShut {NoStop}%
\bibitem [{\citenamefont {Fischer}\ \emph
  {et~al.}(2023{\natexlab{b}})\citenamefont {Fischer}, \citenamefont {Sigrist},
  \citenamefont {Agterberg},\ and\ \citenamefont {Yanase}}]{Fischer2023}%
  \BibitemOpen
  \bibfield  {author} {\bibinfo {author} {\bibfnamefont {Mark~H.}\ \bibnamefont
  {Fischer}}, \bibinfo {author} {\bibfnamefont {Manfred}\ \bibnamefont
  {Sigrist}}, \bibinfo {author} {\bibfnamefont {Daniel~F.}\ \bibnamefont
  {Agterberg}}, \ and\ \bibinfo {author} {\bibfnamefont {Youichi}\ \bibnamefont
  {Yanase}},\ }\bibfield  {title} {\enquote {\bibinfo {title}
  {Superconductivity and local inversion-symmetry breaking},}\ }\href {\doibase
  https://doi.org/10.1146/annurev-conmatphys-040521-042511} {\bibfield
  {journal} {\bibinfo  {journal} {Annu. Rev. Condens. Matter Phys.}\ }\textbf
  {\bibinfo {volume} {14}},\ \bibinfo {pages} {153--172} (\bibinfo {year}
  {2023}{\natexlab{b}})}\BibitemShut {NoStop}%
\bibitem [{Note1()}]{Note1}%
  \BibitemOpen
  \bibinfo {note} {Following Ref.~\protect \citenum {Tsuei2000}, ``dirty''
  junctions refer to those with interface orientation disorder.}\BibitemShut
  {Stop}%
\bibitem [{\citenamefont {Walker}\ and\ \citenamefont
  {Luettmer-Strathmann}(1996)}]{WLS1996}%
  \BibitemOpen
  \bibfield  {author} {\bibinfo {author} {\bibfnamefont {M.~B.}\ \bibnamefont
  {Walker}}\ and\ \bibinfo {author} {\bibfnamefont {J.}~\bibnamefont
  {Luettmer-Strathmann}},\ }\bibfield  {title} {\enquote {\bibinfo {title}
  {Josephson tunneling in high-{${\mathit{T}}_{\mathit{c}}$}
  superconductors},}\ }\href {\doibase 10.1103/PhysRevB.54.588} {\bibfield
  {journal} {\bibinfo  {journal} {Phys. Rev. B}\ }\textbf {\bibinfo {volume}
  {54}},\ \bibinfo {pages} {588--601} (\bibinfo {year} {1996})}\BibitemShut
  {NoStop}%
\bibitem [{\citenamefont {Ambegaokar}\ and\ \citenamefont
  {Baratoff}(1963{\natexlab{a}})}]{Ambegaokar1963}%
  \BibitemOpen
  \bibfield  {author} {\bibinfo {author} {\bibfnamefont {Vinay}\ \bibnamefont
  {Ambegaokar}}\ and\ \bibinfo {author} {\bibfnamefont {Alexis}\ \bibnamefont
  {Baratoff}},\ }\bibfield  {title} {\enquote {\bibinfo {title} {Tunneling
  between superconductors},}\ }\href {\doibase 10.1103/PhysRevLett.10.486}
  {\bibfield  {journal} {\bibinfo  {journal} {Phys. Rev. Lett.}\ }\textbf
  {\bibinfo {volume} {10}},\ \bibinfo {pages} {486--489} (\bibinfo {year}
  {1963}{\natexlab{a}})}\BibitemShut {NoStop}%
\bibitem [{\citenamefont {Ambegaokar}\ and\ \citenamefont
  {Baratoff}(1963{\natexlab{b}})}]{Ambegaokar1963Erratum}%
  \BibitemOpen
  \bibfield  {author} {\bibinfo {author} {\bibfnamefont {Vinay}\ \bibnamefont
  {Ambegaokar}}\ and\ \bibinfo {author} {\bibfnamefont {Alexis}\ \bibnamefont
  {Baratoff}},\ }\bibfield  {title} {\enquote {\bibinfo {title} {Tunneling
  between superconductors},}\ }\href {\doibase 10.1103/PhysRevLett.11.104}
  {\bibfield  {journal} {\bibinfo  {journal} {Phys. Rev. Lett.}\ }\textbf
  {\bibinfo {volume} {11}},\ \bibinfo {pages} {104--104} (\bibinfo {year}
  {1963}{\natexlab{b}})}\BibitemShut {NoStop}%
\bibitem [{\citenamefont {Tanaka}\ and\ \citenamefont
  {Kashiwaya}(1996)}]{Tanaka1996}%
  \BibitemOpen
  \bibfield  {author} {\bibinfo {author} {\bibfnamefont {Yukio}\ \bibnamefont
  {Tanaka}}\ and\ \bibinfo {author} {\bibfnamefont {Satoshi}\ \bibnamefont
  {Kashiwaya}},\ }\bibfield  {title} {\enquote {\bibinfo {title} {Theory of the
  josephson effect in $d$-wave superconductors},}\ }\href {\doibase
  10.1103/PhysRevB.53.R11957} {\bibfield  {journal} {\bibinfo  {journal} {Phys.
  Rev. B}\ }\textbf {\bibinfo {volume} {53}},\ \bibinfo {pages}
  {R11957--R11960} (\bibinfo {year} {1996})}\BibitemShut {NoStop}%
\bibitem [{\citenamefont {Tanaka}\ and\ \citenamefont
  {Kashiwaya}(1997)}]{Tanaka1997}%
  \BibitemOpen
  \bibfield  {author} {\bibinfo {author} {\bibfnamefont {Yukio}\ \bibnamefont
  {Tanaka}}\ and\ \bibinfo {author} {\bibfnamefont {Satoshi}\ \bibnamefont
  {Kashiwaya}},\ }\bibfield  {title} {\enquote {\bibinfo {title} {Theory of
  josephson effects in anisotropic superconductors},}\ }\href {\doibase
  10.1103/PhysRevB.56.892} {\bibfield  {journal} {\bibinfo  {journal} {Phys.
  Rev. B}\ }\textbf {\bibinfo {volume} {56}},\ \bibinfo {pages} {892--912}
  (\bibinfo {year} {1997})}\BibitemShut {NoStop}%
\bibitem [{\citenamefont {Kashiwaya}\ and\ \citenamefont
  {Tanaka}(2000)}]{Kashiwaya2000}%
  \BibitemOpen
  \bibfield  {author} {\bibinfo {author} {\bibfnamefont {Satoshi}\ \bibnamefont
  {Kashiwaya}}\ and\ \bibinfo {author} {\bibfnamefont {Yukio}\ \bibnamefont
  {Tanaka}},\ }\bibfield  {title} {\enquote {\bibinfo {title} {Tunnelling
  effects on surface bound states in unconventional superconductors},}\ }\href
  {\doibase 10.1088/0034-4885/63/10/202} {\bibfield  {journal} {\bibinfo
  {journal} {Reports on Progress in Physics}\ }\textbf {\bibinfo {volume}
  {63}},\ \bibinfo {pages} {1641} (\bibinfo {year} {2000})}\BibitemShut
  {NoStop}%
\bibitem [{\citenamefont {Tanaka}\ \emph {et~al.}(2024)\citenamefont {Tanaka},
  \citenamefont {Tamura},\ and\ \citenamefont {Cayao}}]{Tanaka2024}%
  \BibitemOpen
  \bibfield  {author} {\bibinfo {author} {\bibfnamefont {Yukio}\ \bibnamefont
  {Tanaka}}, \bibinfo {author} {\bibfnamefont {Shun}\ \bibnamefont {Tamura}}, \
  and\ \bibinfo {author} {\bibfnamefont {Jorge}\ \bibnamefont {Cayao}},\
  }\bibfield  {title} {\enquote {\bibinfo {title} {Theory of majorana zero
  modes in unconventional superconductors},}\ }\href {\doibase
  10.1093/ptep/ptae065} {\bibfield  {journal} {\bibinfo  {journal} {Progress of
  Theoretical and Experimental Physics}\ }\textbf {\bibinfo {volume} {2024}},\
  \bibinfo {pages} {08C105} (\bibinfo {year} {2024})}\BibitemShut {NoStop}%
\bibitem [{\citenamefont {Zheng}\ and\ \citenamefont
  {Margine}(2017)}]{Zheng2017}%
  \BibitemOpen
  \bibfield  {author} {\bibinfo {author} {\bibfnamefont {Jing-Jing}\
  \bibnamefont {Zheng}}\ and\ \bibinfo {author} {\bibfnamefont {E.~R.}\
  \bibnamefont {Margine}},\ }\bibfield  {title} {\enquote {\bibinfo {title}
  {Electron-phonon coupling and pairing mechanism in
  {$\ensuremath{\beta}\ensuremath{-}{\mathrm{Bi}}_{2}\mathrm{Pd}$}
  centrosymmetric superconductor},}\ }\href {\doibase
  10.1103/PhysRevB.95.014512} {\bibfield  {journal} {\bibinfo  {journal} {Phys.
  Rev. B}\ }\textbf {\bibinfo {volume} {95}},\ \bibinfo {pages} {014512}
  (\bibinfo {year} {2017})}\BibitemShut {NoStop}%
\bibitem [{\citenamefont {Saib}\ \emph {et~al.}(2017)\citenamefont {Saib},
  \citenamefont {Karaca}, \citenamefont {Tütüncü},\ and\ \citenamefont
  {Srivastava}}]{Saib2017}%
  \BibitemOpen
  \bibfield  {author} {\bibinfo {author} {\bibfnamefont {S.}~\bibnamefont
  {Saib}}, \bibinfo {author} {\bibfnamefont {Ertuǧrul}\ \bibnamefont
  {Karaca}}, \bibinfo {author} {\bibfnamefont {H.M.}\ \bibnamefont
  {Tütüncü}}, \ and\ \bibinfo {author} {\bibfnamefont {G.P.}\ \bibnamefont
  {Srivastava}},\ }\bibfield  {title} {\enquote {\bibinfo {title}
  {Electron-phonon interaction and superconductivity in the multiband
  superconductor {$\beta$-Bi$_2$Pd}},}\ }\href {\doibase
  https://doi.org/10.1016/j.intermet.2017.01.009} {\bibfield  {journal}
  {\bibinfo  {journal} {Intermetallics}\ }\textbf {\bibinfo {volume} {84}},\
  \bibinfo {pages} {136--141} (\bibinfo {year} {2017})}\BibitemShut {NoStop}%
\end{thebibliography}%

\onecolumngrid

\newpage

\setcounter{equation}{0}
\setcounter{section}{0}
\setcounter{figure}{0}
\setcounter{table}{0}
\setcounter{page}{1}
\makeatletter
\renewcommand{\theequation}{S\arabic{equation}}
\renewcommand{\thefigure}{S\arabic{figure}}
\renewcommand{\thesection}{S\arabic{section}}

\renewcommand{\theequation}{S\arabic{equation}}
\renewcommand{\figurename}{FIG.}

\begin{center}
    \textbf{\large{Supplemental Material for ``Spontaneous $\pi$ flux trapping in granular rings of unconventional superconductors''}}
    \\
    \vspace{1em}
    Junyi Zhang and Yi Li
    \\
    \textit{\small Department of Physics and Astronomy, Johns Hopkins University, Baltimore, Maryland 21218, USA}
\end{center}


\section{Microscopic Derivation of the Josephson Coupling Free Energy}
\label{sec:SI_AB}
We derive the first-order Josephson coupling using a microscopic model, 
extending the approach of Ambegaokar and Baratoff~\cite{Ambegaokar1963,Ambegaokar1963Erratum} to multiband superconductors with more complex pairing functions. 
This generalization demonstrates that, 
for weakly coupled superconductors, 
the Josephson free energy takes the form of a sesquilinear functional of the pairing functions, as expressed in Eq.~\eqref{eq:JC_JCFreeEnergy_Sesquilinear} of the main text.

Consider a Josephson junction formed by two superconductors in contact, allowing electrons to tunnel between them. 
The system is described by the Hamiltonian
\begin{equation}
\label{eq:SI_AB_SysHam}
\begin{split}
\mathcal{H} = & \mathcal{H}_{\text{SC}1} + \mathcal{H}_{\text{SC}2} + \mathcal{H}_{T},\\
\mathcal{H}_{\text{SC}n} =& \sum_{\mathbf{k}} \sum_{\mu,\nu} 
\left[
c_{n,\mathbf{k},\mu}^\dagger \hat{H}^a_{\mu,\nu}(\mathbf{k}) c_{n,\mathbf{k},\nu}
+ \left(c_{n,\mathbf{k},\mu}^\dagger \hat{\Delta}^n_{\mu,\nu}(\mathbf{k}) c_{n,-\mathbf{k},\nu}^\dagger + \text{h.c.}
\right) \right],\\
\mathcal{H}_{T}=& 
\mathcal{T} + \mathcal{T}^\dagger 
= \sum_{\mathbf{k}_1,\mathbf{k}_2} 
\sum_{m,n} c_{2,\mathbf{k}_2,m}^\dagger \hat{T}_{m,n}(\mathbf{k}_2,\mathbf{k}_1) c_{1,\mathbf{k}_1,n}+ \text{h.c.},
\end{split}
\end{equation}
where $\mathcal{H}_{\text{SC}n}$ ($n=1,2$) describes the two superconductors on either side of the junction 
[see Fig.~\ref{fig:JJ}(a) of the main text], 
and $\mathcal{H}_T$ encodes the tunneling of electrons across the interface.
The indices $\mu$ and $\nu$ label internal degrees of freedom, 
including spin, orbital, and sublattice components, extending beyond the spin only case considered in Refs.~\citenum{Ambegaokar1963,Ambegaokar1963Erratum}. 
For convenience, we refer to them as ``band indices'', 
although the single-particle Hamiltonian $\hat{H}^a{\mu,\nu}$ need not be diagonal in this basis.

Since the two superconductors are weakly coupled, we treat the tunneling Hamiltonian $\mathcal{H}_T$ as a perturbation.
Using linear response theory, the expectation value of the tunneling current to first order in $\mathcal{H}_T$ is given by
\begin{equation}
\label{eq:SI_AB_TunnlingCurrentLinRespTheory}
\begin{split}
I^{(1)} (t)
=& -\mathrm{i} \int \mathrm{d}t' \theta(t-t')
\langle[I_{T}(t),\mathcal{H}_T(t') ] \rangle,
\end{split}
\end{equation}
where $I_{T} = 
-\mathrm{i} e (\mathcal{T}-\mathcal{T}^\dagger)$ is the tunneling current operator,
and $\theta(t-t')$ is the Heaviside function ensuring causality.
Substituting $\mathcal{H}_T = \mathcal{T} + \mathcal{T}^\dagger$ into Eq.~\eqref{eq:SI_AB_TunnlingCurrentLinRespTheory}, the commutator expands into contributions from different tunneling processes
\begin{equation}
\label{eq:SI_AB_TunnlingCurrentINIJ}
\begin{split}
&I^{(1)} (t)
= I_{N} + I_{J},\\
&I_{N} = - e \int \mathrm{d}t' \theta(t-t')
\Big\{
\langle[\mathcal{T} (t),\mathcal{T}^\dagger(t') ] \rangle
-\langle[\mathcal{T}^\dagger (t),\mathcal{T}(t') ] \rangle
\Big\}\\
&I_{J} = -e \int \mathrm{d}t' \theta(t-t')
\Big\{
\langle[\mathcal{T} (t),\mathcal{T}(t') ] \rangle
-\langle[\mathcal{T}^\dagger (t),\mathcal{T}^\dagger(t') ] \rangle\Big\},
\end{split}
\end{equation}
where $I_N$ corresponds to normal tunneling, 
while $I_J$ describes anomalous pair tunneling and gives rise to the Josephson current.
Defining the retarded anomalous response function
\begin{equation}
\label{eq:SI_AB_RetardedGFDef}
\begin{split}
X^R(t)
= -\mathrm{i} e \int \mathrm{d}t' \theta(t-t')
\langle[\mathcal{T} (t),\mathcal{T}(t') ] \rangle,
\end{split}
\end{equation}
we can calculate the Josephson current as
$I_J= 2 \text{Im} X^R$.

It is more convenient to evaluate the associated Matsubara Green's function $\mathcal{X}(\tau)$
using Wick's theorem
in terms of the fermionic Green's functions
as
\begin{equation}
\label{eq:SI_AB_MatsubaraWickThm}
\begin{split}
&\mathcal{X}(\tau)
= - e \int \mathrm{d}\tau' 
\langle\mathscr{T}_{\tau} \{\mathcal{T} (\tau) \mathcal{T}(\tau') \}\rangle \\
=&- e \int \mathrm{d}\tau' 
\sum_{\substack{\mathbf{k}_1,\mathbf{k}_2\\
\mathbf{k}'_1,\mathbf{k}'_2}} 
\sum_{\substack{\mu,\nu\\
\mu',\nu'}} 
\langle\mathscr{T}_{\tau} \{ 
c_{2,\mathbf{k}_2,\mu}^\dagger \hat{T}_{\mu,\nu}(\mathbf{k}_2,\mathbf{k}_1) c_{1,\mathbf{k}_1,\nu}(\tau)
c_{2,\mathbf{k}'_2,\mu'}^\dagger \hat{T}_{\mu',\nu'}(\mathbf{k}'_2,\mathbf{k}'_1) c_{1,\mathbf{k}'_1,\nu'}(\tau')
\}\rangle \\
=& - e \int \mathrm{d}\tau' 
\sum_{\mathbf{k}_1,\mathbf{k}_2} 
\sum_{\substack{\mu,\nu\\
\mu',\nu'}} 
\left[ 
\hat{T}_{\mu,\nu}(\mathbf{k}_2,\mathbf{k}_1) 
\mathcal{F}_{1,\mathbf{k}_1;\nu',\nu}(\tau',\tau)
\hat{T}_{\mu',\nu'}(-\mathbf{k}_2,-\mathbf{k}_1)
\mathcal{F}_{2,\mathbf{k}_2;\mu',\mu}^*(\tau',\tau)
\right],
\end{split}
\end{equation}
where $\mathcal{F}_{n,\mathbf{k};\mu,\nu}(\tau,\tau')=- \langle \mathscr{T}_{\tau} \{c_{n,-\mathbf{k},\mu} (\tau) c_{n,\mathbf{k},\nu} (\tau') \} \rangle$, $n=1,2$ 
are Gor'kov's anomalous Green's functions.
For the first-order Josephson coupling, 
we expand the anomalous Green’s function 
$\mathcal{F}_n$ 
to leading order in the pairing function
as 
$\mathcal{F}_n \approx g_n^p \hat{\Delta}_n \mathrm{e}^{\mathrm{i}\phi_n} g_n^h$,
where $g_n^{p,h}$ 
are the normal-state propagators for the particle and hole sectors of the Bogoliubov–de Gennes Hamiltonian. 
It follows directly that the anomalous response function
$\mathcal{X}$ is
linear in $\hat{\Delta}_1\mathrm{e}^{\mathrm{i}\phi_1}$
and antilinear in $\hat{\Delta}_2\mathrm{e}^{\mathrm{i}\phi_2}$,
making it a sesquilinear functional of the pairing functions.
Factorizing out the overall $U(1)$ phase dependence, 
we express the response function and the resulting Josephson current as
\begin{equation}
\label{eq:SI_AB_CurrentFunctional}
\begin{split}
X^R
=& -2e \mathcal{I}[\hat{\Delta}_1(\mathbf{k}),\hat{\Delta}_2(\mathbf{k})]\mathrm{e}^{\mathrm{i}(\phi_1-\phi_2)},\\
I_{J} =& 2\mathrm{i} e  
\mathcal{I}[\hat{\Delta}_1(\mathbf{k}),\hat{\Delta}_2(\mathbf{k})]
\mathrm{e}^{\mathrm{i}(\phi_1-\phi_2)} + \text{h.c.},
\end{split}
\end{equation}
where the prefactor of $2$ is introduced for normalization convenience.
This expression clearly shows that the Josephson current depends on the 
gauge invariant phase difference
$\Delta \phi = (\phi_1-\phi_2)$ across the junction.

In Ginzburg–Landau theory, the free energy associated with the first-order Josephson coupling takes the form
\begin{equation}
\label{eq:SI_AB_GinzburgLandauFreeEnergyJosephsonCoupling}
\begin{split}
F_J (\Delta \phi)
=& \mathcal{J}\mathrm{e}^{\mathrm{i}\Delta \phi}+ \text{h.c.}
\end{split}
\end{equation}
The corresponding Josephson current follows from the derivative of the free energy
\begin{equation}
\label{eq:SI_AB_JosephsonCurrentPhaseAndersonFormula}
\begin{split}
I_J =& 2e \frac{\partial F_J}{\partial \Delta \phi}, 
=2 \mathrm{i} e \mathcal{J}\mathrm{e}^{\mathrm{i}\Delta \phi }+ \text{h.c.}.
\end{split}
\end{equation}
Comparing Eqs.\eqref{eq:SI_AB_CurrentFunctional}
and~\eqref{eq:SI_AB_JosephsonCurrentPhaseAndersonFormula},
we identify the Josephson coupling coefficient as $\mathcal{J} = \mathcal{I}[\hat{\Delta}_1(\mathbf{k}),\hat{\Delta}_2(\mathbf{k})]$.
This provides a microscopic foundation for the Ginzburg–Landau form of the Josephson free energy given in Eq.~\eqref{eq:JC_JCFreeEnergy_Sesquilinear} of the main text.

\section{Angular Dependence of Josephson Couplings for $d$- and $p$-Wave Superconductors}
\label{sec:SI_JCpdSC}
The angular dependence of the Josephson coupling, 
dictated by Eq.~\eqref{eq:JC_AngularFormFactor} 
of the main text, 
follows from symmetry considerations.
In this section, we derive this dependence explicitly for $d$- and $p$-wave superconductors.
In particular, 
when applied to $d$-wave superconductors such as the cuprates, 
it reproduces the generalized Sigrist–Rice formula, 
along with a symmetry-allowed correction term to the original Sigrist–Rice formula~\cite{Sigrist1992}  that has been previously identified by Walker and Luettmer-Strathmann using a series expansion method~\cite{WLS1996}.

\subsection{Single-Band $d$-Wave Superconductors}
\label{sec:SI_JCpdSC_dSC}
For $d$-wave superconductors with in-plane rotational symmetries, such as the cuprates, it is convenient to choose a basis $\mathcal{B}_d = \{ \hat{\Delta}_{d_{x^2-y^2}},\hat{\Delta}_{d_{xy}} \}$ that spans the space of a two-dimensional representation.
Notably, these two basis functions have opposite mirror parities under the mirror reflection $\mathcal{M}_y$,
i.e.,
$\mathcal{M}_y\hat{\Delta}_{d_{x^2-y^2}} = +\hat{\Delta}_{d_{x^2-y^2}} $
and $\mathcal{M}_y\hat{\Delta}_{d_{xy}} = -\hat{\Delta}_{d_{xy}} $.
Using the sesquilinearity of the functional 
$\mathcal{I}$ we find
\begin{equation}
\label{eq:SI_JCpdSC_dWaveMirrorParityForbidden}
\begin{split}
&\mathcal{I}[\hat{\Delta}_{d_{x^2-y^2}}, \hat{\Delta}_{d_{xy}}]
=\mathcal{I}[\mathcal{M}_y\hat{\Delta}_{d_{x^2-y^2}}, \mathcal{M}_y\hat{\Delta}_{d_{xy}}]
=-\mathcal{I}[\hat{\Delta}_{d_{x^2-y^2}}, \hat{\Delta}_{d_{xy}}].
\end{split}
\end{equation}
As a result, the inter-channel Josephson coupling 
$\mathcal{I}[\hat{\Delta}_{d_{x^2-y^2}}, \hat{\Delta}_{d_{xy}}]=0$.
By the same argument,
$\mathcal{I}[ \hat{\Delta}_{d_{xy}},\hat{\Delta}_{d_{x^2-y^2}}]$
also vanishes.
In contrast, the intra-channel couplings are allowed.
Therefore, 
we summarize the values of $\mathcal{I}$ 
evaluated on the basis $\mathcal{B}_d$ as
\begin{equation}
\label{eq:SI_JCpdSC_dWaveCouplingBasis}
\begin{split}
&\mathcal{I}[\hat{\Delta}_{d_{x^2-y^2}}, \hat{\Delta}_{d_{x^2-y^2}}]
=-I_v,\\
&\mathcal{I}[\hat{\Delta}_{d_{x^2-y^2}}, \hat{\Delta}_{d_{xy}}]
=0,\\
&\mathcal{I}[\hat{\Delta}_{d_{xy}}, \hat{\Delta}_{d_{x^2-y^2}}]
=0,\\
&\mathcal{I}[\hat{\Delta}_{d_{xy}}, \hat{\Delta}_{d_{xy}}]
=-I_d,
\end{split}
\end{equation}
where $I_v$ and $I_d$ are positive constants
The minus signs are fixed such that the free energy is minimized 
when the two grains are aligned and the relative phase difference vanishes.

We now derive the angular dependence of the Josephson coupling by examining 
how the pairing functions transform under in-plane rotations. 
Consider a rotation by an angle $\alpha$, 
denoted $R_{\alpha}$.
Under this transformation, 
the pairing functions rotate within the basis $\mathcal{B}_d$ as
\begin{equation}
\label{eq:SI_JCpdSC_dWaveRotation}
\begin{split}
&\hat{\Delta}^{\alpha}_{d_{x^2-y^2}} 
= \cos(2\alpha)\hat{\Delta}_{d_{x^2-y^2}} + \sin(2\alpha)\hat{\Delta}_{d_{xy}},\\
&\hat{\Delta}^{\alpha}_{d_{xy}} 
= -\sin(2\alpha)\hat{\Delta}_{d_{x^2-y^2}} + \cos(2\alpha)\hat{\Delta}_{d_{xy}}.
\end{split}
\end{equation}
Substituting these rotated pairing functions into the sesquilinear functional $\mathcal{I}$, 
we obtain the angular dependence of the Josephson coupling amplitudes between two $d$-wave grains with orientations $\alpha_1$ and $\alpha_2$ as
\begin{equation}
\label{eq:SI_JCpdSC_dWaveJCAngularDep}
\begin{split}
\mathcal{I}[\hat{\Delta}^{\alpha_1}_{d_{x^2-y^2}}, \hat{\Delta}^{\alpha_2}_{d_{x^2-y^2}}]
=&  \cos(2\alpha_1)\cos(2\alpha_2)
\mathcal{I}[\hat{\Delta}_{d_{x^2-y^2}}, \hat{\Delta}_{d_{x^2-y^2}}]\\
&+ \cos(2\alpha_1)\sin(2\alpha_2)
\mathcal{I}[\hat{\Delta}_{d_{x^2-y^2}}, \hat{\Delta}_{d_{xy}}]\\
&+ \sin(2\alpha_1)\cos(2\alpha_2)
\mathcal{I}[\hat{\Delta}_{d_{xy}}, \hat{\Delta}_{d_{x^2-y^2}}]\\
&+\sin(2\alpha_1)\sin(2\alpha_2)
\mathcal{I}[\hat{\Delta}_{d_{xy}}, \hat{\Delta}_{d_{xy}}]\\
=&-\cos(2\alpha_1)\cos(2\alpha_2)I_v
-\sin(2\alpha_1)\sin(2\alpha_2)I_d,\\
\mathcal{I}[\hat{\Delta}^{\alpha_1}_{d_{xy}}, \hat{\Delta}^{\alpha_2}_{d_{xy}}]
=& -\sin(2\alpha_1)\sin(2\alpha_2)I_v
-\cos(2\alpha_1)\cos(2\alpha_2)I_d.
\end{split}
\end{equation}
Therefore, the Josephson coupling free energy between two $d$-wave  grains,
such as those found in cuprates,
takes the form
\begin{equation}
\label{eq:SI_JCpdSC_GeneralizedSRForumla}
\begin{split}
F_J[\hat{\Delta}^{\alpha_1}_{d_{x^2-y^2}}, \hat{\Delta}^{\alpha_2}_{d_{x^2-y^2}}]
=&- 2\left[\cos(2\alpha_1)\cos(2\alpha_2)I_v
+\sin(2\alpha_1)\sin(2\alpha_2)I_d
\right] \cos (\phi_1 - \phi_2).
\end{split}
\end{equation}
When $I_d = 0$, 
Eq.~\eqref{eq:SI_JCpdSC_GeneralizedSRForumla} reduces to 
$F_J =- 2\cos(2\alpha_1)\cos(2\alpha_2)I_v \cos (\phi_1 - \phi_2)$,
which is also known as the Sigrist–Rice formula~\cite{Sigrist1992}.
This expression has been used in the design of tricrystal experiments for probing unconventional pairing in clean Josephson junctions~\cite{Tsuei1994,Tsuei2000}.
The second term proportional to $I_d$ is also symmetry-allowed, 
but it was not included in the original discussion by Sigrist and Rice.  
It was later identified by Walker and Luettmer-Strathmann using a series expansion approach~\cite{WLS1996}.
Our symmetry-based derivation naturally recovers this full expression, 
providing a generalized form of the Sigrist–Rice formula given in Eq.~\eqref{eq:SI_JCpdSC_GeneralizedSRForumla}.

For chiral $d$-wave superconductors, 
the pairing functions can be expressed as
\begin{equation}
\label{eq:SI_JCpdSC_CdWavePairingFunction}
\begin{split}
\hat{\Delta}_{d_{\pm 2}} 
=& \hat{\Delta}_{d_{x^2-y^2}} \pm \mathrm{i} \hat{\Delta}_{d_{xy}},
\end{split}
\end{equation}
and the corresponding Josephson coupling amplitudes are given by
\begin{equation}
\label{eq:SI_JCpdSC_CdWaveJC}
\begin{split}
\mathcal{I}[\hat{\Delta}_{d_{\pm2}}, \hat{\Delta}_{d_{\pm2}}]
=& - (I_v + I_d),\\
\mathcal{I}[\hat{\Delta}_{d_{\pm2}}, \hat{\Delta}_{d_{\mp2}}]
=& - (I_v - I_d),
\end{split}
\end{equation}
where $I_v$ and $I_d$ are the same constants introduced earlier.
Under an in-plane rotation $R_{\alpha}$,
the pairing functions transform as
\begin{equation}
\label{eq:SI_JCpdSC_CdWaveRotation}
\begin{split}
\hat{\Delta}^{\alpha}_{d_{\pm 2}} 
=& \mathrm{e}^{\mp 2\mathrm{i} \alpha} \hat{\Delta}_{d_{\pm 2}},
\end{split}
\end{equation}
leading to the following angular dependence of the Josephson couplings
\begin{equation}
\label{eq:SI_JCpdSC_CdWaveJCAngularDep}
\begin{split}
\mathcal{I}[\hat{\Delta}^{\alpha_1}_{d_{\pm2}}, \hat{\Delta}^{\alpha_2}_{d_{\pm2}}]
=&  \mathrm{e}^{\mp 2\mathrm{i} \alpha_1} 
\mathrm{e}^{\pm 2\mathrm{i} \alpha_2} 
\mathcal{I}[\hat{\Delta}_{d_{\pm2}}, \hat{\Delta}_{d_{\pm2}}]
=  -\mathrm{e}^{\mp 2\mathrm{i} (\alpha_1-\alpha_2)} 
(I_v+I_d ),\\
\mathcal{I}[\hat{\Delta}^{\alpha_1}_{d_{\pm2}}, \hat{\Delta}^{\alpha_2}_{d_{\mp2}}]
=&  \mathrm{e}^{\mp 2\mathrm{i} \alpha_1} 
\mathrm{e}^{\mp 2\mathrm{i} \alpha_2} 
\mathcal{I}[\hat{\Delta}_{d_{\pm2}},
\hat{\Delta}_{d_{\mp2}}]
= -\mathrm{e}^{\mp 2\mathrm{i} (\alpha_1+\alpha_2)} 
(I_v-I_d),
\end{split}
\end{equation}
where we have used the sesquilinearity of the functional $\mathcal{I}$.
The Josephson coupling free energy between two chiral $d$-wave grains of the same chirality then takes the form
\begin{equation}
\label{eq:SI_JCpdSC_FreeEnergyCdWave}
\begin{split}
F_J[\hat{\Delta}^{\alpha_1}_{d_{\pm2}}, \hat{\Delta}^{\alpha_2}_{d_{\pm2}}]
=&- 2(I_v +I_d) \cos\left[
\mp 2 (\alpha_1 - \alpha_2)
+ (\phi_1 - \phi_2)
\right].
\end{split}
\end{equation}

\subsection{Single-Band $p$-Wave Superconductors}
\label{sec:SI_JCpdSC_pSC}
For $p$-wave superconductors, we analyze the Josephson couplings in parallel with the $d$-wave case.
we analyze the Josephson couplings for $p$-wave superconductors.
For simplicity, we first neglect spin-orbit coupling 
and assume that the spin part of the Cooper pairs transforms trivially under in-plane rotational symmetries.
It is then convenient to adopt the mirror eigenbasis
$\mathcal{B}_p = \{ \hat{\Delta}_{p_{x}} = p_x (\mathrm{i}\sigma_y) \sigma_z,\hat{\Delta}_{p_{y}} = p_y (\mathrm{i}\sigma_y) \sigma_z \}$
which diagonalizes the mirror reflection $\mathcal{M}_y$.
In this basis, the Josephson coupling functional evaluates to
\begin{equation}
\label{eq:SI_JCpdSC_pWaveCouplingBasis}
\begin{split}
&\mathcal{I}[\hat{\Delta}_{p_{x}},\hat{\Delta}_{p_{x}}]
=-I_x,\\
&\mathcal{I}[\hat{\Delta}_{p_{x}},\hat{\Delta}_{p_{y}}]
=0,\\
&\mathcal{I}[\hat{\Delta}_{p_{y}},\hat{\Delta}_{p_{x}}]
=0,\\
&\mathcal{I}[\hat{\Delta}_{p_{y}},\hat{\Delta}_{p_{y}}]
=-I_y.
\end{split}
\end{equation}
The inter-channel Josephson couplings vanish due to opposite mirror parities under $\mathcal{M}_y$,
while the intra-channel couplings
$-I_x$ and $-I_y$ 
are symmetry allowed.
Under an in-plane rotation $R_{\alpha}$,
the pairing functions transform as
\begin{equation}
\label{eq:SI_JCpdSC_pWaveRotation}
\begin{split}
&\hat{\Delta}^{\alpha}_{p_{x}} 
= \cos(\alpha)\hat{\Delta}_{p_{x}} + \sin(\alpha)\hat{\Delta}_{p_{y}},\\
&\hat{\Delta}^{\alpha}_{p_{y}} 
= -\sin(\alpha)\hat{\Delta}_{p_{x}} + \cos(\alpha)\hat{\Delta}_{p_{y}}.
\end{split}
\end{equation}
Substituting these into the sesquilinear functional $\mathcal{I}$,
we obtain the angular dependence of the Josephson coupling amplitudes
\begin{equation}
\label{eq:SI_JCpdSC_pWaveJCAngularDep}
\begin{split}
\mathcal{I}[\hat{\Delta}^{\alpha_1}_{p_{x}}, \hat{\Delta}^{\alpha_2}_{p_{x}}]
=&-\cos(\alpha_1)\cos(\alpha_2)I_x
-\sin(\alpha_1)\sin(\alpha_2)I_y,\\
\mathcal{I}[\hat{\Delta}^{\alpha_1}_{p_{y}}, \hat{\Delta}^{\alpha_2}_{p_{y}}]
=&-\sin(\alpha_1)\sin(\alpha_2)I_x
-\cos(\alpha_1)\cos(\alpha_2)I_y,
\end{split}
\end{equation}
Moreover, the Josephson coupling free energy between two $p$-wave grains with the same orbital channel follows as
\begin{equation}
\label{eq:SI_JCpdSC_pWaveGeneralizedSRForumla}
\begin{split}
F_J[\hat{\Delta}^{\alpha_1}_{p_{x}}, \hat{\Delta}^{\alpha_2}_{p_{x}}]
=&- 2\left[\cos(\alpha_1)\cos(\alpha_2)I_x
+\sin(\alpha_1)\sin(\alpha_2)I_y
\right] \cos (\phi_1 - \phi_2),\\
F_J[\hat{\Delta}^{\alpha_1}_{p_{y}}, \hat{\Delta}^{\alpha_2}_{p_{y}}]
=&- 2\left[\sin(\alpha_1)\sin(\alpha_2)I_x
+\cos(\alpha_1)\cos(\alpha_2)I_y
\right] \cos (\phi_1 - \phi_2).
\end{split}
\end{equation}

For chiral $p$-wave superconductors, 
the pairing functions can be expressed as
\begin{equation}
\label{eq:SI_JCpdSC_CpWavePairingFunction}
\begin{split}
\hat{\Delta}_{p_{\pm}} 
=& \hat{\Delta}_{p_{x}} \pm \mathrm{i} \hat{\Delta}_{p_{y}}.
\end{split}
\end{equation}
and the corresponding Josephson coupling amplitudes are given by
\begin{equation}
\label{eq:App_JCpdDisorder_CdWaveJC}
\begin{split}
\mathcal{I}[\hat{\Delta}_{p_{\pm}}, \hat{\Delta}_{p_{\pm}}]
=& - (I_x + I_y),\\
\mathcal{I}[\hat{\Delta}_{p_{\pm}}, \hat{\Delta}_{p_{\mp}}]
=& - (I_x - I_y).
\end{split}
\end{equation}
Under an in-plane rotation $R_{\alpha}$,
the pairing functions transform as
\begin{equation}
\label{eq:SI_JCpdSC_CpWaveRotation}
\begin{split}
\hat{\Delta}^{\alpha}_{p_{\pm }} 
=& \mathrm{e}^{\mp \mathrm{i} \alpha} \hat{\Delta}_{p_{\pm }},
\end{split}
\end{equation}
leading to the following angular dependence of the Josephson couplings
\begin{equation}
\label{eq:SI_JCpdSC_CpWaveJCAngularDep}
\begin{split}
\mathcal{I}[\hat{\Delta}^{\alpha_1}_{p_{\pm}}, \hat{\Delta}^{\alpha_2}_{p_{\pm}}]
=&  -\mathrm{e}^{\mp \mathrm{i} (\alpha_1-\alpha_2)} 
(I_x+I_y ),\\
\mathcal{I}[\hat{\Delta}^{\alpha_1}_{p_{\pm}}, \hat{\Delta}^{\alpha_2}_{p_{\mp}}]
=& -\mathrm{e}^{\mp \mathrm{i} (\alpha_1+\alpha_2)} 
(I_x-I_y).
\end{split}
\end{equation}
The Josephson coupling free energy between two chiral $p$-wave grains of the same chirality then takes the form
\begin{equation}
\label{eq:SI_JCpdSC_FreeEnergyCpWave}
\begin{split}
F_J[\hat{\Delta}^{\alpha_1}_{p_{\pm}}, \hat{\Delta}^{\alpha_2}_{p_{\pm}}]
=&- 2(I_x +I_y) \cos\left[
\mp (\alpha_1 - \alpha_2)
+ (\phi_1 - \phi_2)
\right].
\end{split}
\end{equation}

In the presence of spin-orbit coupling, 
it is more natural to work in the chiral basis
$\mathcal{B}_{p_{\pm j_z}}$,
where the Cooper pairs carry definite total angular momentum due to in-plane rotation symmetries.
Although the explicit form of the chiral pairing functions can be complicated,
it is their transformation under in-plane rotation $R_{\alpha}$
\begin{equation}
\label{eq:SI_JCpdSC_CpWaveRotationJz}
\begin{split}
\hat{\Delta}^{\alpha}_{p_{\pm j_z}} 
=& \mathrm{e}^{\mp j_z \mathrm{i} \alpha} \hat{\Delta}_{p_{\pm j_z}},
\end{split}
\end{equation}
that governs the angular dependence of the Josephson couplings.
Using this symmetry property, 
the Josephson coupling free energy between two chiral $p$-wave grains of the same chirality is given by
\begin{equation}
\label{eq:SI_JCpdSC_FreeEnergyCpWaveJz}
\begin{split}
F_J[\hat{\Delta}^{\alpha_1}_{p_{\pm j_z}}, \hat{\Delta}^{\alpha_2}_{p_{\pm j_z}}]
=&- 2I_{p_{\pm j_z}} \cos\left[
\mp j_z(\alpha_1 - \alpha_2)
+ (\phi_1 - \phi_2)
\right].
\end{split}
\end{equation}

\section{Effects of Interface Orientation Disorder}
\label{sec:SI_Disorder}
Interface disorder is unavoidable in realistic experimental settings 
and can substantially alter the angular dependence of Josephson couplings, 
as noted by Tsuei and collaborators~\cite{Tsuei1994}.
In the context of $d$-wave superconductors such as the cuprates, 
they proposed a modified version of the Sigrist–Rice formula for junctions in the  ``dirty limit'', 
where the disorder effect at the junction interface is strong.
To avoid confusion with the conventional use of ``dirty'' for bulk disorders, 
we adopt the term dirty (Josephson) junction to describe junctions with significant interface disorders, 
and dirty junction limit for strongly disordered cases.
Interestingly, the modified Sigrist–Rice formula proposed for the dirty junction limit in cuprates 
coincides formally with the clean-junction result derived by Walker and Luettmer-Strathmann~\cite{WLS1996,Tsuei2000} when one sets $I_v = -I_d$ in Eq.~\eqref{eq:SI_JCpdSC_dWaveJCAngularDep}. However, this agreement appears to be purely coincidental from a mathematical standpoint and lacks a physical justification grounded in symmetry principles.

Here, we examine the effects of interface disorder arising from fluctuations in the interface orientation, 
following the physical mechanism identified by Tsuei and collaborators~\cite{Tsuei1994}. 
We model the disorder as a random distribution of local interface orientations, enabling a more systematic and quantitative analysis.
For $d$-wave superconductors, 
we find that the modification proposed by Tsuei et al.~\cite{Tsuei1994} remains accurate,
while that for $p$-wave superconductors contains correction terms of order $30\%$, 
which may qualitatively alter the conclusions. 
Interestingly, we also find that Josephson couplings in chiral superconductors are comparatively robust against such interface orientation disorder.

As illustrated in Fig.~\ref{fig:JJ} of the main text, 
interface orientation disorder can be modeled as local random deviations $\beta$ from the average junction orientation. 
The effect of these deviations on the Josephson couplings can be incorporated by modifying the grain orientations as $\alpha_n \rightarrow \tilde{\alpha}_n = \alpha_n - \beta$. 
The total Josephson coupling is then obtained by averaging over the disorder variable $\beta$. 
Assuming $\beta$ follows a probability distribution $P(\beta)$, the disorder-averaged Josephson coupling is given by
\begin{equation}
\label{eq:SI_JCpdDisorder_InterfaceDisorderAvgDef}
\begin{split}
\bar{\mathcal{I}}
=&\overline{\mathcal{I}[\hat{\Delta}^{\tilde{\alpha}_1}_L, 
\hat{\Delta}^{\tilde{\alpha}_2}_R]}_{\beta}
=\int \mathrm{d} \beta P(\beta) \mathcal{I}[\hat{\Delta}^{\alpha_1-\beta}_L, 
\hat{\Delta}^{\alpha_2-\beta}_R].
\end{split}
\end{equation} 
We now show that the disorder-averaged weighting factors are governed by the generating function $\Pi(l)$ associated with the probability distribution $P(\beta)$. 
For simplicity and without loss of generality, we assume that the orientation fluctuation $\beta$ follows a normal distribution, $\beta \sim \mathcal{N}(0, \sigma_\beta^2)$. 
The corresponding generating function takes the form
\begin{equation}
\label{eq:SI_JCpdDisorder_GeneratingFunctionGaussian}
\begin{split}
\Pi(l) 
=& \int \mathrm{d} \beta P(\beta) \mathrm{e}^{\mathrm{i} l \beta}
=\frac{1}{\sqrt{2\pi \sigma_\beta^2}} \int \mathrm{d} \beta \mathrm{e}^{-\frac{\beta^2}{2\sigma_\beta^2}} \mathrm{e}^{\mathrm{i} l \beta}
=\mathrm{e}^{- \frac{1}{2}\sigma^2_{\beta} l^2}.
\end{split}
\end{equation}

Applying this framework to $d$-wave superconductors, 
whose Josephson coupling is governed by the generalized Sigrist–Rice formula [Eq.~\eqref{eq:SI_JCpdSC_GeneralizedSRForumla}], 
and using the generating function in Eq.~\eqref{eq:SI_JCpdDisorder_GeneratingFunctionGaussian}, 
we find that the angular dependent factors in the square brackets are averaged to
\begin{equation}
\label{eq:SI_JCpdDisorder_AngularFormFactorDisorderAvgdWave}
\begin{split}
\overline{\cos(2\tilde{\alpha}_1)\cos(2\tilde{\alpha}_2)}_{\beta}
=& \frac{1}{2}\cos[2(\alpha_1-\alpha_2)] w^d_- 
+ \frac{1}{2}\cos[2(\alpha_1+\alpha_2)]
w^d_+,\\
\overline{\sin(2\tilde{\alpha}_1)\sin(2\tilde{\alpha}_2)}_{\beta}
=& \frac{1}{2} \cos[2(\alpha_1-\alpha_2)] w^d_-
- \frac{1}{2}\cos[2(\alpha_1+\alpha_2)]
w^d_+,
\end{split}
\end{equation}
where the weighting factors are 
$w^d_- = \Pi(0) = 1$ 
and $w^d_+ = \Pi(4) = \mathrm{e}^{-8\sigma^2_{\beta}}$.
In the dirty junction limit with large $\sigma_\beta$, 
Tsuei et al. estimated a maximal disorder $\sigma_\beta \approx \pi/4$ for tetragonal lattices~\cite{Tsuei1994}, 
leading to
$w_+^d \approx \exp(-\pi^2/2) \approx 7.2\times 10^{-3}$.
which implies that the $\cos[2(\alpha_1 + \alpha_2)]$ term is exponentially suppressed by over two orders of magnitude. 
As a result, both disorder-averaged prefactors reduce to a form proportional to $\cos[2(\alpha_1 - \alpha_2)]$. 
Thus, in the dirty junction limit, the Josephson coupling simplifies to
\begin{equation}
\label{eq:SI_JCpdDisorder_GeneralizedSRForumlaDisorder}
\begin{split}
\bar{\mathcal{I}} \approx& - \frac{1}{2}  (I_v +I_d) \cos[2(\alpha_1-\alpha_2)],\\
\bar{F}_J
\approx&- (I_v+I_d)\cos[2(\alpha_1-\alpha_2)]\cos (\phi_1 - \phi_2).
\end{split}
\end{equation}
which coincides with the form proposed by Tsuei and collaborators~\cite{Tsuei1994,Tsuei2000}.

We emphasize, however, the distinction between this result and the clean-junction expression [Eq.~\eqref{eq:SI_JCpdSC_GeneralizedSRForumla}]. 
While the same $\cos[2(\alpha_1 - \alpha_2)]$ dependence can be recovered in the clean limit by setting $I_v = -I_d$, 
as noted in Refs.~\citenum{WLS1996,Tsuei2000}, 
such a choice lacks physical justification: no symmetry enforces equal magnitudes and opposite signs for these two couplings. 
The agreement is therefore a mathematical coincidence rather than a symmetry-enforced equivalence.
In contrast, Eq.~\eqref{eq:SI_JCpdDisorder_GeneralizedSRForumlaDisorder} arises robustly from the disorder averaging procedure and remains valid regardless of the specific values of $I_v$ and $I_d$, 
as long as $w_+^d$ is sufficiently suppressed,
which provides a more rigorous and physically grounded justification for the angular dependence of Josephson couplings in the presence of interface disorder.

In cuprates with tetragonal lattice symmetry, 
the relative magnitudes of $I_v$ and $I_d$ can be qualitatively estimated as follows. 
When $\alpha_1 = \alpha_2 = 0$, 
the Josephson coupling is entirely governed by $I_v$, 
corresponding to tunneling along the crystallographic axes. 
In the thin-wall limit, electron tunneling occurs predominantly through nearest-neighbor bonds, 
and from Eq.~\eqref{eq:SI_AB_MatsubaraWickThm}, 
we estimate $I_v \sim |t|^2$, 
where $t$ denotes the nearest-neighbor hopping amplitude.
In contrast, when $\alpha_1 = \alpha_2 = \pi/4$, 
the coupling is governed solely by $I_d$, which arises from tunneling along the diagonal direction. 
In this configuration, 
the tunneling is mediated by next-nearest-neighbor hopping processes, 
yielding $I_d \sim |t'|^2$, 
where $t'$ is the next-nearest-neighbor hopping amplitude.
Since typically $|t'/t| \sim 0.1$ in cuprates, 
we estimate the ratio $I_d/I_v \approx |t'|^2/|t|^2 \sim 10^{-2}$, indicating that $I_d \ll I_v$.

We now turn to the Josephson couplings in a dirty Josephson junction composed of $p$-wave superconductors. In parallel with the $d$-wave case, the angular prefactors in Eq.~\eqref{eq:SI_JCpdSC_pWaveGeneralizedSRForumla} are modified by the interface disorder and become
\begin{equation}
\label{eq:SI_JCpdDisorder_pWaveAngularFormFactorDisorderAvgdWave}
\begin{split}
&\overline{\cos(\tilde{\alpha}_1)\cos(\tilde{\alpha}_2)}_{\beta}
= \frac{1}{2}\cos(\alpha_1-\alpha_2) w^p_- 
+ \frac{1}{2}\cos(\alpha_1+\alpha_2)
w^p_+,\\
&\overline{\sin(\tilde{\alpha}_1)\sin(\tilde{\alpha}_2)}_{\beta}
= \frac{1}{2} \cos(\alpha_1-\alpha_2) w^p_-
- \frac{1}{2}\cos(\alpha_1+\alpha_2)
w^p_+,
\end{split}
\end{equation}
where the weight factors are given by $w^p_- = \Pi(0) = 1$ and $w^p_+ = \Pi(2) = \exp(-2\sigma_\beta^2)$.
Substituting these into Eq.~\eqref{eq:SI_JCpdSC_pWaveGeneralizedSRForumla}, the disorder-averaged Josephson coupling becomes
\begin{equation}
\label{eq:SI_JCpdDisorder_pWaveGeneralizedSRForumlaDisorder}
\begin{split}
&\bar{\mathcal{I}} =
- \frac{1}{2} \cos(\alpha_1-\alpha_2)(I_v +I_d) w_-^p
- \frac{1}{2} \cos(\alpha_1+\alpha_2)(I_v -I_d) w_+^p,\\
&\bar{F}_J =
- \left[ \cos(\alpha_1-\alpha_2)(I_v +I_d) w_-^p
+ \cos(\alpha_1+\alpha_2)(I_v -I_d) w_+^p \right]\cos (\phi_1 - \phi_2).
\end{split}
\end{equation}

For a tetragonal lattice in the dirty junction limit, 
we estimate $\sigma_\beta \approx \pi/4$, 
which yields $\exp(-2\sigma_\beta^2) \approx 0.29$. 
In contrast to the $d$-wave case, 
the $\cos(\alpha_1 + \alpha_2)$ term remains appreciable in $p$-wave superconductors. 
Despite being exponentially suppressed, its weight factor still retains roughly $30\%$ of that of the $\cos(\alpha_1 - \alpha_2)$ term, 
highlighting a key distinction between the two cases. 
As a result, applying the approximated form proposed by Tsuei and collaborators to junctions composed of $p$-wave  superconductors can lead to substantial quantitative deviations and potentially incorrect conclusions.

For chiral superconductors, the effects of interface disorder manifest differently. The angular dependence of the disorder-averaged Josephson couplings for chiral $d$- and $p$-wave superconductors, given in Eqs.~\eqref{eq:SI_JCpdSC_CdWaveJCAngularDep} and \eqref{eq:SI_JCpdSC_CpWaveJCAngularDep}, can be evaluated as
\begin{equation}
\label{eq:SI_JCpdDisorder_CSCJCDisorder}
\begin{split}
\overline{\mathcal{I}[\hat{\Delta}^{\tilde{\alpha}_1}_{\pm m},
\hat{\Delta}^{\tilde{\alpha}_2}_{\pm m}]}_{\beta}
=& \mathcal{I}[\hat{\Delta}^{\alpha_1}_{\pm m},
\hat{\Delta}^{\alpha_2}_{\pm m}],\\
\overline{\mathcal{I}
[\hat{\Delta}^{\tilde{\alpha}_1}_{\pm m}, \hat{\Delta}^{\tilde{\alpha}_2}_{\mp m}]}_{\beta}
=& w'_c\mathcal{I}
[\hat{\Delta}^{\alpha_1}_{\pm m}, \hat{\Delta}^{\alpha_2}_{\mp m}],
\end{split}
\end{equation}
where $w'_c = \Pi(2m) = \mathrm{e}^{-2m^2\sigma^2{\beta}}$. 
This result shows that Josephson couplings between chiral superconductors of the same chirality are unaffected by interface orientation disorder, 
whereas couplings between opposite chiralities are exponentially suppressed.

\section{Phase Diagrams of Three-Grain Superconducting Rings}
\label{sec:SI_3GR}

\begin{figure}[htb]
\includegraphics[width=0.7\linewidth]{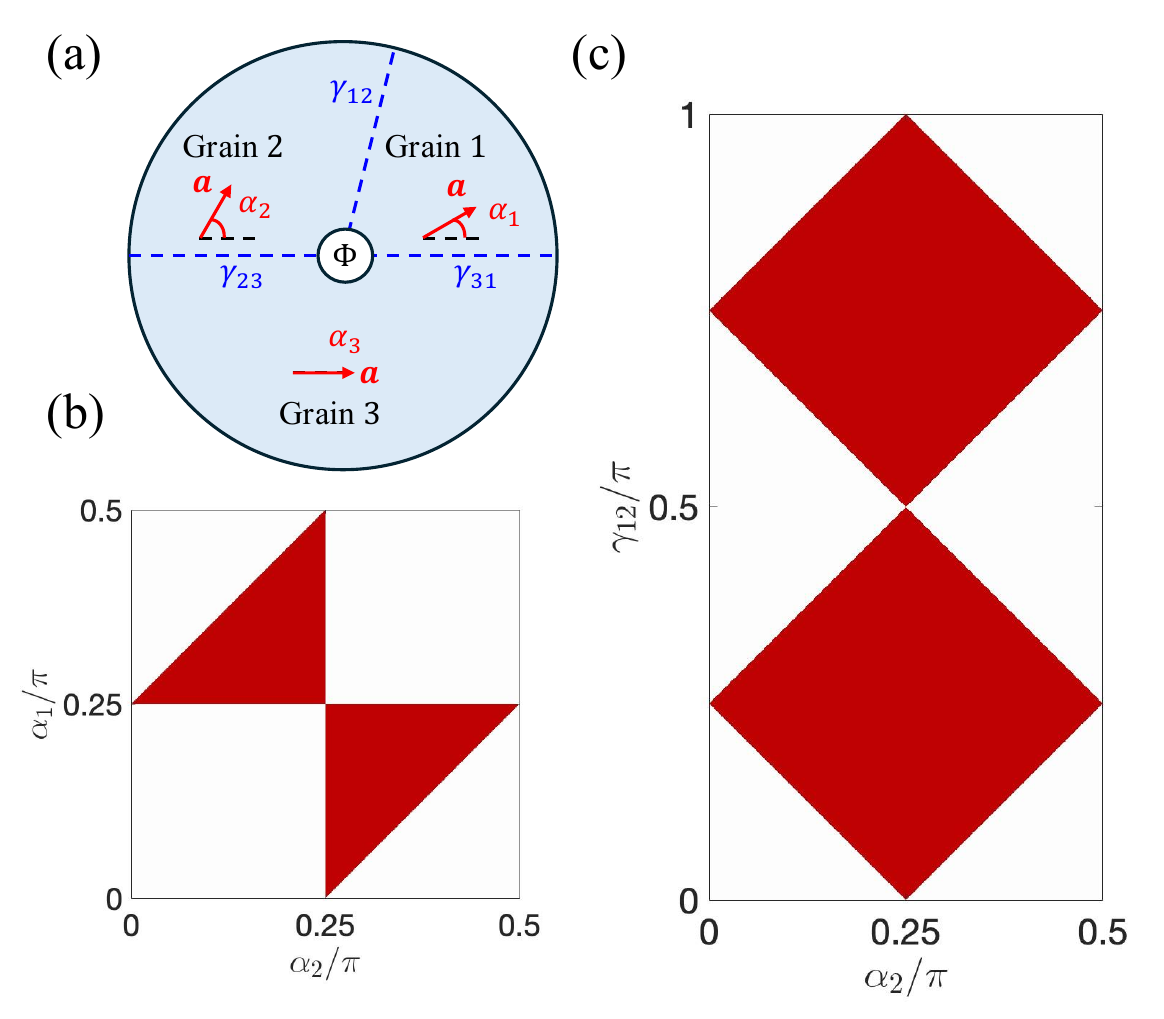}
\caption{\label{fig:SFig_3GR_dWavePhaseDiagrams} 
Geometric configuration and phase diagrams of a three-grain ring composed of$d$-wave superconductors.
(a). Schematic of a three-grain ring, showing grain orientations $\{\alpha_{1,2,3}\}$,
and interface orientations $\{\gamma_{12,23,31}\}$.
For simplicity, $\alpha_3 =0$.
(b). Phase diagram in the dirty junction limit, plotted as a function of $\alpha_1$ and $\alpha_2$.
The red-shaded region denotes grain configurations that support spontaneous $\pi$-flux trapping.
(c). Phase diagram for clean Josephson junctions, plotted as a function of $\alpha_{12}$ and $\alpha_2$
under the constraints $\gamma_{23} = \pi$, $\gamma_{31} = 2\pi$, and the relation $\alpha_1+\alpha_2 = \pi/2$.
The red-shaded region again indicates 
$\pi$-ring configurations.
}
\end{figure}

To facilitate a direct comparison between the phase diagrams supporting $\pi$-flux trapping in three-grain rings of $d$-wave superconductors, 
as studied by Tsuei and collaborators~\cite{Tsuei2000}, 
and those of helical equal-spin pairing discussed in the main text, 
we present a detailed derivation of the Josephson coupling energies in three-grain superconducting rings, both in the clean and dirty junction limits.

We begin by reproducing the phase diagram of three-grain rings composed of $d$-wave superconductors, 
originally studied by Tsuei et al.~\cite{Tsuei2000}. 
The three-grain ring (tricrystal) geometry is illustrated in SFig.~\ref{fig:SFig_3GR_dWavePhaseDiagrams}
(a), 
following the setup considered in Refs.~\citenum{Tsuei1994,Tsuei2000}. 
In their notation, the grain boundary angles are defined with respect to the interface between grains 1 and 3. 
For consistency and ease of comparison, 
we note that our angle $\gamma_{12}$ corresponds to the angle labeled ``$\beta$'' in Ref.~\citenum{Tsuei2000}.

Assuming that the magnetic flux is confined to the central hole of the superconducting ring, 
we adopt the gauge
\begin{equation}
\label{eq:SI_3GR_GaugePotential}
\begin{split}
\mathbf{A}(\mathbf{r}) =&  \frac{\Phi_B}{2\pi r} \mathbf{e}_{\theta},
\end{split}
\end{equation}
where $\Phi_B$ denotes the total magnetic flux,
and $\mathbf{e}_{\theta}$ is the unit vector in the azimuthal direction.
The associated $U(1)$ phases acquired by superconducting grains due to the magnetic flux are given by
\begin{equation}
\label{eq:SI_3GR_GaugeFieldU1PhaseSliding}
\begin{split}
\Phi(\mathbf{r}) 
=& \frac{e^*}{\hbar} \int^{\mathbf{r}} \mathrm{d}\mathbf{r} \cdot \mathbf{A}
= \theta (\Phi_B/ \phi_0),
\end{split}
\end{equation}
where $\theta$ is the polar angle of the position $\mathbf{r}$,
and $\phi_0$ is the superconducting flux quantum.
Using the generalized Sigrist–Rice formula for clean Josephson junctions [Eq.~\eqref{eq:SI_JCpdSC_GeneralizedSRForumla}],
the total Josephson coupling free energy of the three-grain ring is given by
\begin{equation}
\label{eq:SI_3GR_FreeEnergy3GR}
\begin{split}
F_J =& - \Big\{
I_v \cos [2(\alpha_1-\tilde{\gamma}_{12})] \cos [2(\alpha_2-\tilde{\gamma}_{12})]\\
&+I_d \sin [2(\alpha_1-\tilde{\gamma}_{12})] \sin [2(\alpha_2-\tilde{\gamma}_{12})]
\Big\}
\cos(\tilde{\phi}_1-\tilde{\phi}_2)\\
&- \Big\{
I_v \cos [2(\alpha_2-\tilde{\gamma}_{23})] \cos [2(\alpha_3-\tilde{\gamma}_{23})]\\
&+I_d \sin [2(\alpha_2-\tilde{\gamma}_{23})] \sin [2(\alpha_3-\tilde{\gamma}_{23})]
\Big\}
\cos(\tilde{\phi}_2-\tilde{\phi}_3)\\
&- \Big\{
I_v \cos [2(\alpha_3-\tilde{\gamma}_{31})] \cos [2(\alpha_1-\tilde{\gamma}_{31})]\\
&+I_d \sin [2(\alpha_3-\tilde{\gamma}_{31})] \sin [2(\alpha_1-\tilde{\gamma}_{31})]
\Big\}
\cos(\tilde{\phi}_3-\tilde{\phi}_1 + \Phi),
\end{split}
\end{equation}
where 
$\tilde{\gamma}_{ij}= \gamma_{ij} -\pi/2$
accounts for the difference in convention used to  define the interface orientation compared to Ref.~\citenum{Tsuei2000}, 
and $\Phi = 2\pi\Phi_B/\phi_0$ 
is the accumulated total $U(1)$ phase around the ring.
Due to our choice of branch cut for the polar angle along the grain boundary between grain 1 and grain 3, this phase $\Phi$ appears only in the last term.

Even in this relatively simple three-grain geometry, 
several tunable parameters exist, including the grain orientations
$\{\alpha_{i}\}$
and the grain boundary interface orientations $\{\gamma_{ij}\}$.
To simplify the analysis, Tsuei and collaborators~\cite{Tsuei2000} considered the specific configuration
$\gamma_{31}=0$, $\gamma_{23}=\pi$, and $\alpha_3=0$,
together with the constraint 
$\alpha_1+\alpha_2 = \pi/2$.
In addition, they adopted the original Sigrist–Rice formula, 
which is mathematically equivalent to setting $I_d = 0$ in Eq.~\eqref{eq:SI_3GR_FreeEnergy3GR}.
Under these assumptions, the total free energy simplifies to
\begin{equation}
\label{eq:SI_3GR_FreeEnergy3GRSRClean}
\begin{split}
F_J =& - 
I_v \cos (\pi-2\alpha_2-2\gamma_{12}) \cos (2\alpha_2-2\gamma_{12})
\cos(\tilde{\phi}_1-\tilde{\phi}_2)\\
&-
I_v \cos (2\alpha_2)
\cos(\tilde{\phi}_2-\tilde{\phi}_3)\\
&- 
I_v \cos (\pi-2\alpha_2)
\cos(\tilde{\phi}_3-\tilde{\phi}_1 + \Phi).
\end{split}
\end{equation}
Noting that the angular factors in the last two lines of Eq.~\eqref{eq:SI_3GR_FreeEnergy3GRSRClean} always appear with opposite signs, 
the condition for spontaneously trapping a $\pi$-flux requires
$\cos (\pi-2\alpha_2-2\gamma_{12}) \cos (2\alpha_1-2\gamma_{12})>0$,
or, equivalently, 
$\cos (2\alpha_2+2\gamma_{12}) \cos (2\alpha_2-2\gamma_{12})<0$, 
which reproduces the criterion obtained by Tsuei and collaborators~\cite{Tsuei2000}.
The phase boundaries are then given by the conditions
$\alpha_2 \pm \gamma_{12} = \pi/4 + k\pi/2, k\in \mathbb{Z}$.
SFig.~\ref{fig:SFig_3GR_dWavePhaseDiagrams} (c) shows the phase diagram of a three-grain ring composed of clean Josephson junctions between $d$-wave superconductors.
The red-shaded region indicates the grain configurations that support spontaneous $\pi$-flux trapping.

In the dirty junction limit, the effective Josephson coupling is given by Eq.~\eqref{eq:SI_JCpdDisorder_GeneralizedSRForumlaDisorder}, 
and the free energy of the three-grain ring takes the form
\begin{equation}
\label{eq:SI_3GR_FreeEnergy3GRSRDirty}
\begin{split}
F_J =& - 
I^{12} \cos (2\alpha_1-2 \alpha_2) 
\cos(\tilde{\phi}_1-\tilde{\phi}_2)\\
&-
I^{23} \cos (2\alpha_2-2 \alpha_3) 
\cos(\tilde{\phi}_2-\tilde{\phi}_3)\\
&- 
I^{31} \cos (2\alpha_3-2 \alpha_1) 
\cos(\tilde{\phi}_3-\tilde{\phi}_1 + \Phi).
\end{split}
\end{equation}
In this limit, the free energy no longer depends on the interface orientations $\{\gamma_{ij}\}$,
but only on the relative orientations of the superconducting grains.
To simplify the analysis, we follow Ref.~\citenum{Tsuei2000} and fix $\alpha_3 =0$.
The phase boundaries for $\pi$-ring formation are then determined by the conditions $\alpha_1,\alpha_2,\alpha_1-\alpha_2  = \pi/4 + k\pi/2, k\in \mathbb{Z}$.
SFig.~\ref{fig:SFig_3GR_dWavePhaseDiagrams}(b) presents the corresponding phase diagram for a tricrystal ring composed of dirty Josephson junctions between $d$-wave superconductors.
The red-shaded region indicates the configurations that support spontaneous $\pi$-flux trapping.

Fig.~\ref{fig:3GR} 
(b) in the main text presents the phase diagram of a three-grain ring composed of helical equal-spin pairing superconductors. 
It closely resembles the phase diagram shown in SFig.~\ref{fig:SFig_3GR_dWavePhaseDiagrams} (b) for $d$-wave superconductors in the dirty junction limit. 
This similarity originates from the analogous angular dependence of the Josephson coupling free energies in the two cases, 
given by Eq.~\eqref{eq:SI_3GR_FreeEnergy3GRSRClean} and Eq.~\eqref{eq:HESP_HESPFreeEnergy} of the main text, respectively.
Since Eq.~\eqref{eq:HESP_HESPFreeEnergy} remains valid for helical ESP superconducting rings irrespective of the degree of interface disorder, 
the $\pi$-ring test in a tricrystal geometry is expected to be significantly more robust against orientation fluctuations compared to the $d$-wave case.

\section{$\pi$-Ring Probability}
\label{sec:SI_PiRing}

Although Tsuei and collaborators~\cite{Tsuei1994} demonstrated $\pi$-flux trapping in a tricrystal experiment using YBCO, 
fabricating granular superconducting rings with precise control over both grain and interface orientations remains experimentally challenging.
In practice, it is more feasible to pattern polycrystalline thin films into ring geometries, where the number of grains $N_g$ can be much larger than three.
As $N_g$ increases, the grain configurations that support $\pi$-ring formation form a complex subset within a high-dimensional parameter space.
Given the lack of precise control over these configurations, 
it is more heuristic to examine statistically the probability of obtaining a $\pi$-ring in systems with $N_g>3$.

The general criterion for realizing a $\pi$-ring with arbitrary grain number $N_g$ is given in Eq.~\eqref{eq:FreeEnergy_FluxTrappingCriterion} of the main text.
In this section, we first compute the probability of obtaining a $\pi$-ring in three- and four-grain rings composed of helical equal-spin pairing superconductors.
We then use Monte Carlo simulations to evaluate the probability of $\pi$-ring formation in rings with $N_g \geq 3$ for both helical equal-spin pairing and $d$-wave superconductors, in both clean and dirty junction limits.
Our results show that as $N_g$ increases, the probability $P$ of realizing a $\pi$-ring grows and asymptotically approaches $P = 1/2$ in the large-$N_g$ limit.
This suggests that fabricating granular superconducting rings by patterning polycrystalline thin films can be a viable approach, provided the underlying unconventional superconductor supports such states.

The free energy of Josephson coupling for an $N_g$-grain ring composed of helical equal spin pairing superconductor is given by Eq.~\eqref{eq:HESP_HESPFreeEnergy} of the main text
that we quote here again as
\begin{equation}\label{eq:SI_PiRing_FreeEnergyNgGRHESP}
\begin{split}
F_J =&\sum_{n=1}^{N_g} 
-2I_{n,n+1}\cos (\alpha_{n} - \alpha_{n+1})
\cos\Big[(\tilde{\phi}_{n}-\tilde{\phi}_{n+1}) 
+ (\Phi_{n,+} -\Phi_{n+1,-}) \Big].
\end{split}
\end{equation}
Similar to the case of three-grain $d$-wave superconducting ring as discussed in Sec.~\ref{sec:SI_3GR},
we observe that
there is a translational symmetry in configurational space 
as the free energy does not change if $\alpha_n \rightarrow \alpha'_n = \alpha_n + c_{\alpha}$.

When $N_g = 3$, the Josephson coupling free energy for a ring of helical equal-spin pairing superconductors takes the form
\begin{equation}
\label{eq:SI_PiRing_FreeEnergy3GRHESP}
\begin{split}
F_J =& - 
I^{12} \cos (\alpha_1-\alpha_2) 
\cos(\tilde{\phi}_1-\tilde{\phi}_2)\\
&-
I^{23} \cos (\alpha_2-\alpha_3) 
\cos(\tilde{\phi}_2-\tilde{\phi}_3)\\
&- 
I^{31} \cos (\alpha_3-\alpha_1) 
\cos(\tilde{\phi}_3-\tilde{\phi}_1 + \Phi).
\end{split}
\end{equation}
The boundaries separating $0$- and $\pi$-ring configurations are determined by the conditions
\begin{equation}
\label{eq:SI_PiRing_PhaseBoundary3GRHESP}
\begin{split}
\alpha_1-\alpha_2 = & \frac{\pi}{2} + k_{1,2} \pi,\\
\alpha_2-\alpha_3 = & \frac{\pi}{2} + k_{2,3} \pi,\\
\alpha_3-\alpha_1 = & \frac{\pi}{2} + k_{3,1} \pi,
\end{split}
\end{equation}
where $k_{n,n+1} \in \mathbb{Z}$.
Due to translational symmetry in configuration space, we may, without loss of generality, 
set $\alpha_1 = 0$. 
This simplifies the analysis and allows the phase diagram to be visualized as a function of $(\alpha_2, \alpha_3)$, 
as shown in Fig.~\ref{fig:3GR} 
(b) of the main text.
Assuming the grain orientations $\alpha_n$ are independently and uniformly distributed over $[0, 2\pi]$, 
The probability of forming a $\pi$-ring is proportional to the area of the butterfly-shaped red-shaded region in the phase diagram.
From this, we obtain
\begin{equation}
\label{eq:SI_PiRing_ProbPiRing3GRHESP}
\begin{split}
P(N_g=3)=\frac{1}{4}.
\end{split}
\end{equation}

When $N_g = 4$, the Josephson coupling free energy for a ring composed of helical equal-spin pairing superconductors takes the form
\begin{equation}
\label{eq:eq:SI_PiRing_FreeEnergy4GRHESP}
\begin{split}
F_J =& - 
I^{12} \cos (\alpha_1-\alpha_2) 
\cos(\tilde{\phi}_1-\tilde{\phi}_2)\\
&-
I^{23} \cos (\alpha_2-\alpha_3) 
\cos(\tilde{\phi}_2-\tilde{\phi}_3)\\
&- 
I^{34} \cos (\alpha_3-\alpha_4) 
\cos(\tilde{\phi}_3-\tilde{\phi}_4 )\\
&- 
I^{41} \cos (\alpha_4-\alpha_1) 
\cos(\tilde{\phi}_4-\tilde{\phi}_1+ \Phi).
\end{split}
\end{equation}
The phase boundaries separating $0$- and $\pi$-ring configurations are determined by the conditions
\begin{equation}
\label{eq:SI_PiRing_PhaseBoundary4GRHESP}
\begin{split}
\alpha_1-\alpha_2 = & \frac{\pi}{2} + k_{1,2} \pi,\\
\alpha_2-\alpha_3 = & \frac{\pi}{2} + k_{2,3} \pi,\\
\alpha_3-\alpha_4 = & \frac{\pi}{2} + k_{3,4} \pi,\\
\alpha_4-\alpha_1 = & \frac{\pi}{2} + k_{4,1} \pi,
\end{split}
\end{equation}
where $k_{n,n+1} \in \mathbb{Z}$.
Without loss of generality, we fix $\alpha_1 = 0$, reducing the configuration space to $(\alpha_2, \alpha_3, \alpha_4)$.
The resulting phase boundaries in this reduced space are visualized in 
Fig.~\ref{fig:4GR} 
(b) of the main text.
In particular, when $\alpha_1 = \alpha_4 = 0$,
Eq.~\eqref{eq:SI_PiRing_PhaseBoundary4GRHESP} reduces to Eq.~\eqref{eq:SI_PiRing_PhaseBoundary3GRHESP},
implying that the region of $(\alpha_2, \alpha_3)$ supporting a $\pi$-ring coincides with the butterfly-shaped domain found in the three-grain case.
This region is highlighted in yellow in 
Fig.~\ref{fig:4GR} 
(c).
More generally, representative slices of the phase diagram at fixed $\alpha_4 \in (0, \pi/2)$ and $\alpha_4 \in (\pi/2, \pi)$
are shown in 
Figs.~\ref{fig:4GR} 
(d) and (e), respectively.
In each case, the yellow-shaded regions indicate the configurations that support a $\pi$-ring.

For a general value of $\alpha_4$, the probability of obtaining a $\pi$-ring (proportional to the area of the yellow-shaded regions shown in 
Figs.~\ref{fig:4GR} 
 (d) and 
\ref{fig:4GR} 
 (e)), 
denoted as $p_{2,3}(\alpha_4)$, 
is given by
\begin{equation}
\label{eq:eq:SI_PiRing_Area4GRHESPalpha4}
\begin{split}
&p_{2,3}(\alpha_4) =
\begin{cases} 
\frac{A_-}{\pi^2} = \frac{1}{2\pi^2}\left[\left(\frac{\pi}{2} +\alpha_4 \right)^2 + \left(\frac{\pi}{2} -\alpha_4 \right)^2\right],
&\alpha_4 \in (0,\pi/2),\\
\frac{A_+}{\pi^2} = \frac{1}{2\pi^2}\left[\left(\alpha_4 - \frac{\pi}{2}  \right)^2 + \left(\frac{3\pi}{2} -\alpha_4 \right)^2\right],
&\alpha_4 \in (\pi/2,\pi).
\end{cases}
\end{split}
\end{equation}
Then we obtain total probability of forming a $\pi$-ring in a four-grain ring of helical equal-spin pairing superconductors as
\begin{equation}
\label{eq:eq:SI_PiRing_ProbPiRing4GRHESP}
\begin{split}
P(N_g=4) = \frac{1}{\pi}\int_{0}^{\pi} \mathrm{d} \alpha_4 p_{2,3}(\alpha_4) = \frac{1}{3}.
\end{split}
\end{equation}

\begin{figure}[htb]
\includegraphics[width=0.5\linewidth]{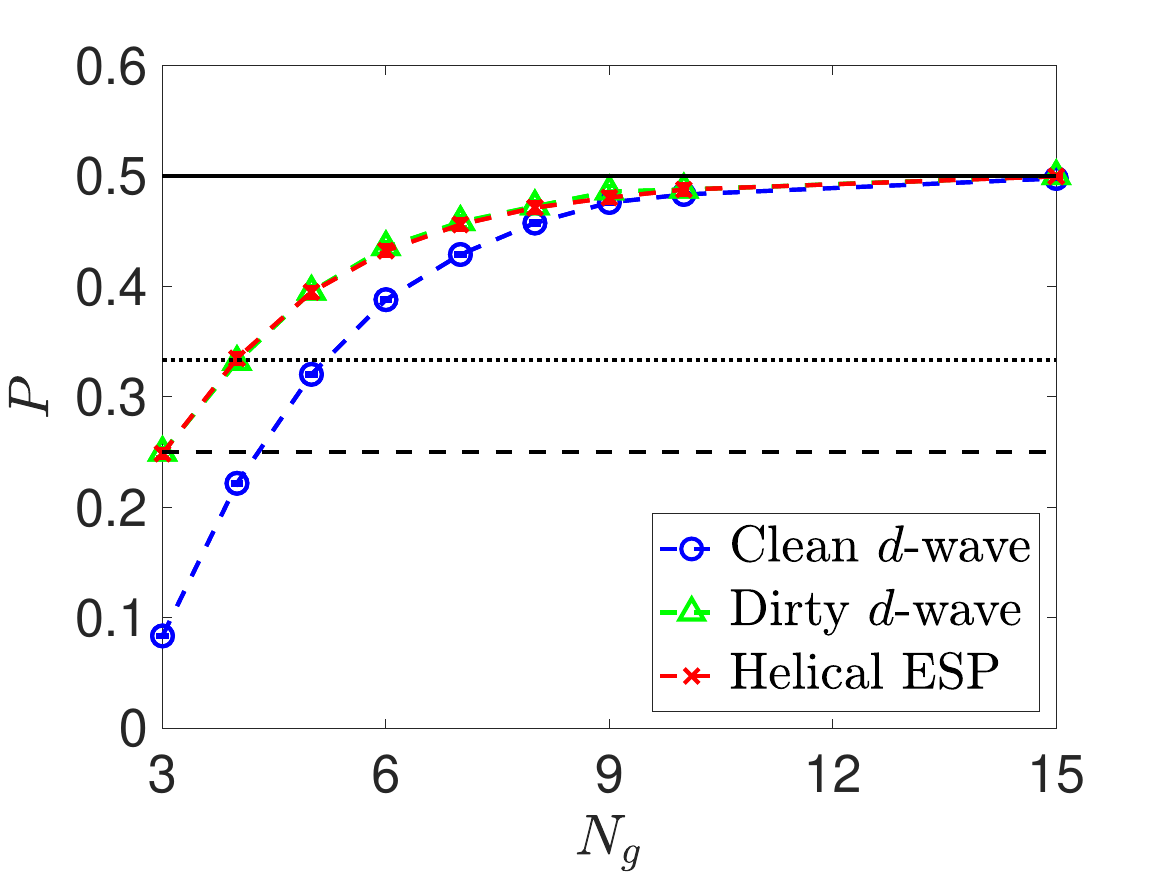}
\caption{\label{fig:SFig_PiRing_PiRingProb}
Probability of forming a 
$\pi$-ring as a function of number of grains, $N_g$.
$P(N_g)$ are presented for
$d$-wave superconductors with clean (blue circles) and dirty (green triangles) Josephson junctions, 
as well as for helical equal-spin pairing superconductors (red crosses).
The dashed (resp. dotted) line indicates the exact value of $P(N_g=3)=1/4$ (resp. $P(N_g=4)=1/4$) for helical equal-spin pairing superconductors.
The solid line indicates $P=1/2$ in the large $N_g$ limit. 
}
\end{figure}

To evaluate the probability $P(N_g)$ of forming a $\pi$-ring in $N_g$-grain superconducting rings, 
we perform Monte Carlo simulations for both helical equal-spin pairing and $d$-wave superconductors under clean and dirty junction limits.
For each $N_g$, we generate $10^5$ 
random geometric configurations, sampling both grain orientations and interface angles uniformly. The resulting probabilities of obtaining a 
$\pi$-ring are presented in SFig.~\ref{fig:SFig_PiRing_PiRingProb}.
Our results show that for small $N_g$,
$d$-wave superconductors with clean Josephson junctions exhibit significantly lower probabilities of $\pi$-ring formation 
compared to ones with dirty junctions 
and helical ESP systems. 
However, as $N_g$ increases, the probability $P(N_g)$ also increases and, in all cases, converges to $P={1}/{2}$ in the $N_g$ limit.

\section{A Minimal Two-Band Model for {$\beta$-$\ch{Bi_2Pd}$}}
\label{sec:SI_M2BM}
\begin{figure}[htb]
\includegraphics[width=0.5\linewidth]{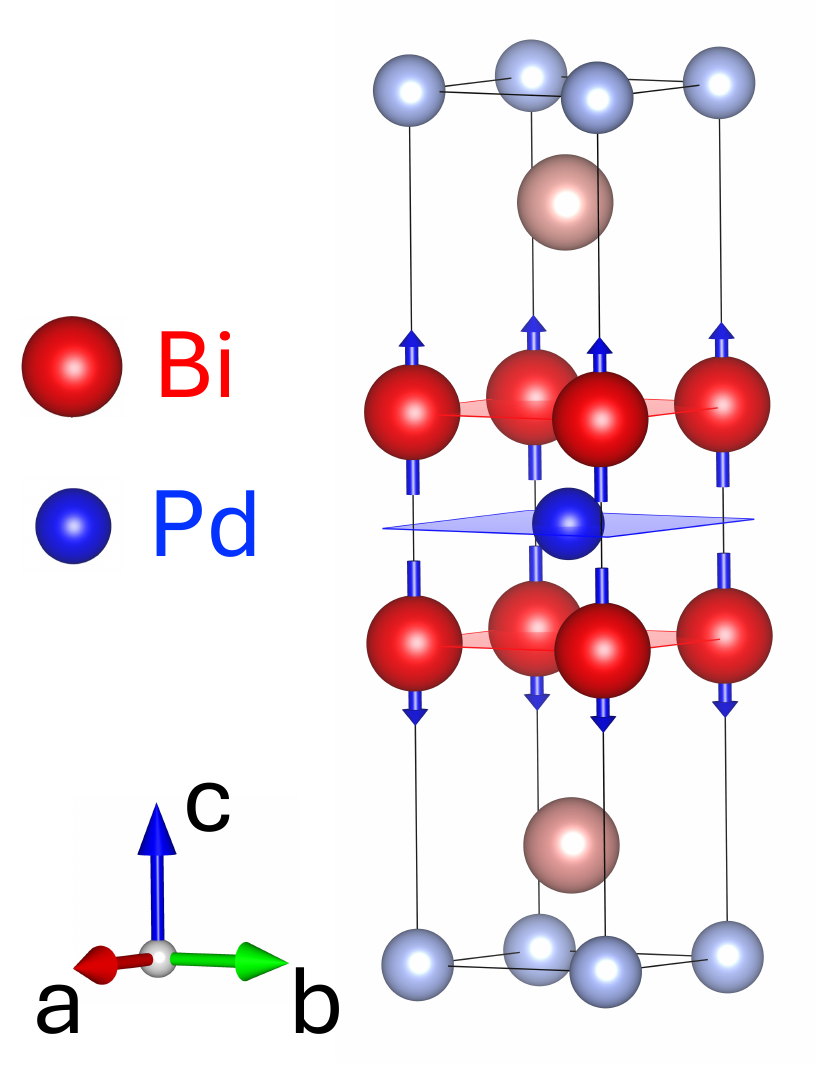}
\caption{\label{fig:SFig_CrystalStructure}
Crystal structure of $\beta$-$\ch{Bi_2Pd}$. $\ch{Bi}$ atoms (red, $4e$ Wyckoff positions) 
and $\ch{Pd}$ atoms (blue, $2a$ Wyckoff positions) 
form $\ch{Bi}$–$\ch{Pd}$–$\ch{Bi}$ trilayers stacked along the $c$-axis, 
with successive trilayers shifted by the body-centered translation $(1/2,1/2,1/2)$.
Within each trilayer, 
$\ch{Bi}$ and $\ch{Pd}$ form square nets in the $ab$-plane, 
with the $\ch{Pd}$ layer (blue plane) lying on the mirror plane $\mathcal{M}_z$ 
and hosting the inversion center at the $\ch{Pd}$ sites, 
whereas the $\ch{Bi}$ layers (red planes) lack mirror symmetry 
and locally break inversion symmetry, 
as indicated by the symmetry-allowed polar vectors (blue arrows).
}
\end{figure}

$\beta$-$\ch{Bi_2Pd}$ crystallizes in the tetragonal space group $I4/mmm$. 
A conventional unit cell is shown in Fig.~\ref{fig:SFig_CrystalStructure}.
The $\ch{Bi}$ atoms (red) occupy the $4e$ Wyckoff positions, 
and the $\ch{Pd}$ atoms (blue) reside at the $2a$ Wyckoff position that coincides with an inversion center of the lattice.
The $\ch{Pd}$ and $\ch{Bi}$ atoms form $\ch{Bi}$–$\ch{Pd}$–$\ch{Bi}$ trilayers stacked along the $c$-axis, 
with successive trilayers shifted by the body-centered translation $(1/2,1/2,1/2)$.
Despite its structural simplicity, the material exhibits rather intricate Fermi surfaces as a result of multiple bands crossing the Fermi level.
The layered structure leads to predominantly quasi-two-dimensional (2D) Fermi surface geometry.
These bands originate mainly from $\ch{Bi}$ $p$-orbitals, 
while the $\ch{Pd}$ $d$-orbitals are fully occupied.
As a result, $\beta$-$\ch{Bi_2Pd}$ does not display strong electron correlation effects or magnetic ordering.

In the presence of both time-reversal and global inversion symmetries, 
each Fermi surface in $\beta$-$\ch{Bi_2Pd}$ remains doubly degenerate.
However, despite this degeneracy, 
the Fermi surfaces exhibit hidden spin polarization with nontrivial spin textures, 
as a consequence of spin-orbit coupling (SOC) arising from \emph{local} inversion symmetry breaking.
Specifically, although the crystal as a whole possesses inversion centers at the $\ch{Pd}$ sites, 
individual $\ch{Bi}$ layers do not, 
as indicated by the symmetry-allowed polar vectors (blue arrows) in Fig.~\ref{fig:SFig_CrystalStructure}.
Under inversion, a given $\ch{Bi}$ layer is mapped to its counterpart on the opposite side of the $\ch{Pd}$ layer, rather than onto itself, and the associated polar vectors reverse their orientation.
As a result, electrons moving within the $\ch{Bi}$ planes experience a locally noncentrosymmetric environment, 
giving rise to a non-vanishing local SOC field.

To model the effects of SOC and local inversion symmetry breaking on superconducting pairing,
we consider a minimal two-band model that incorporates all the essential ingredients.
Specifically, we consider two square-lattice layers arranged in an $AA$ stacking configuration.
For simplicity, we assume that each site hosts a single orbital that transforms as a $\ch{Bi}$ $p_z$-orbital under the space group symmetries.
The band Hamiltonian is given by
\begin{equation}\label{eq:SI_M2BM_BandHam}
\begin{split}
\mathcal{H}_b =& \sum_{\mathbf{k}}
\Big\{
\sum_{l,s}
[\epsilon_0(\mathbf{k})-\mu] c_{\mathbf{k},l,s}^\dagger c_{\mathbf{k},l,s}
+
\sum_{s}
[(t_{\perp})
c_{\mathbf{k},l=1,s}^\dagger c_{\mathbf{k},l=2,s} + \text{h.c.}]\\
&+ 
\sum_{l}\sum_{s,s'}
\mathbf{g}_l(\mathbf{k}) \cdot \bm{\sigma}_{s,s'}
c_{\mathbf{k},l,s}^\dagger c_{\mathbf{k},l,s'}
\Big\},
\end{split}
\end{equation}
where $\epsilon_0(\mathbf{k})$ describes the in-plane dispersion,
$t_\perp$ denotes the interlayer hopping amplitude,
and $\mathbf{g}_l(\mathbf{k})$ encodes the SOC arising from local inversion asymmetry within each layer.
Physically, the absence of mirror symmetry $\mathcal{M}_z$ in each $\ch{Bi}$ layer 
(red planes in Fig.~\ref{fig:SFig_CrystalStructure}) 
allows a local polar field perpendicular to the plane, 
indicated by the blue arrows along $c$-axis in Fig.~\ref{fig:SFig_CrystalStructure}.
Therefore, each $\ch{Bi}$ layer supports a Rashba-type SOC with $\mathbf{g}_l(\mathbf{k}) \propto \mathbf{e}_z \times \mathbf{k}$.
Nevertheless,
since the polar vectors reverse their orientation for adjacent layers connected by the global inversion symmetry of the crystal,
the assign SOC amplitudes 
should also change the sign as 
$\alpha_{l} = (-)^l\alpha_R$, 
leading to the explicit form
$\mathbf{g}_l(\mathbf{k}) = (-1)^l \alpha_R \mathbf{e}_z \times \mathbf{k}$,
where $\alpha_R$ is the strength of the SOC.

The band eigenstates are obtained by diagonalizing the band Hamiltonian $\mathcal{H}_b$.
The resulting energy dispersions are given by
$\epsilon_\pm(\mathbf{k}) = \epsilon_0(\mathbf{k}) - \mu \pm \sqrt{\alpha_R^2 k^2 +(t_{\perp})^2}$,
where $k = |\mathbf{k}|$,
The corresponding eigenstates can be represented in terms of tensor products of spin- and layer-spinors as
\begin{equation}\label{eq:SI_M2BM_BandHamEig}
\begin{split}
\epsilon_+(\mathbf{k}): 
\chi_1=\eta^l_{+,+}\otimes \xi^s_+,\\
\epsilon_-(\mathbf{k}): 
\chi_2=\eta^l_{-,+}\otimes \xi^s_+,\\
\epsilon_-(\mathbf{k}): 
\chi_3=\eta^l_{-,-}\otimes \xi^s_-,\\
\epsilon_+(\mathbf{k}): \chi_4=\eta^l_{+,-}\otimes \xi^s_-,
\end{split}
\end{equation}
where the spin- and layer-spinors are explicitly defined as follows
\begin{equation}\label{eq:SI_M2BM_EigSpinorDef}
\begin{split}
&\xi^s_{+}(\mathbf{k}) =\frac{1}{\sqrt{2}} \begin{pmatrix}
\mathrm{e}^{-\mathrm{i}\theta_{\mathbf{k}}}\\
\mathrm{i}
\end{pmatrix},
\xi^s_{-}(\mathbf{k}) = \frac{1}{\sqrt{2}} \begin{pmatrix}
\mathrm{i}\\
\mathrm{e}^{+\mathrm{i}\theta_{\mathbf{k}}}
\end{pmatrix},\\
&\eta^l_{+,+}(\mathbf{k}) = \begin{pmatrix}
+\cos (\gamma/2)\\ 
+\sin (\gamma/2)
\end{pmatrix},
\eta^l_{-,+}(\mathbf{k}) = \begin{pmatrix}
-\sin (\gamma/2)\\ 
+\cos (\gamma/2)
\end{pmatrix},\\
&\eta^l_{+,-}(\mathbf{k}) = \begin{pmatrix}
+\sin (\gamma/2)\\ 
+\cos (\gamma/2)
\end{pmatrix},
\eta^l_{-,-}(\mathbf{k}) = \begin{pmatrix}
-\cos (\gamma/2)\\ 
+\sin (\gamma/2)
\end{pmatrix}.
\end{split}
\end{equation}
Here, the angle $\theta_{\mathbf{k}}$ denotes the argument of $k_x+\mathrm{i}k_y$,
and the angle $\gamma$ is defined by $\tan\gamma = t_{\perp}/(\alpha_R k)$
characterizing the layer hybridization.
The superscripts $s$ and $l$ indicate that the corresponding spinors are defined in the spin and layer subspaces, respectively.

In this model, 
the time-reversal symmetry is represented by
$\hat{\mathcal{T}}: \psi(\mathbf{k}) \mapsto \mathrm{i} \sigma_2 \psi^*(-\mathbf{k})$,
and the inversion symmetry is represented by
$\hat{\mathcal{I}}: \psi(\mathbf{k}) \mapsto \lambda_1 \psi(-\mathbf{k})$,
where $\sigma_{i}$ and $\lambda_i$ are Pauli matrices acting on the spin and layer degrees of freedom, respectively.
It is straightforward to verify that the degenerate band eigenstates defined in Eq.~\eqref{eq:SI_M2BM_BandHamEig} 
are connected by the combined
$\hat{\mathcal{I}}\circ \hat{\mathcal{T}}$ operation,
i.e.,
$\hat{\mathcal{I}}\circ \hat{\mathcal{T}} [\chi_1(\mathbf{k}) ]= -\chi_4(\mathbf{k})$
and 
$\hat{\mathcal{I}}\circ \hat{\mathcal{T}} [\chi_2(\mathbf{k})] = +\chi_3(\mathbf{k})$.

Experimental evidence indicates that superconductivity in $\beta$-$\ch{Bi_2Pd}$ is nodeless and preserves time-reversal symmetry. In addition, theoretical studies suggest that the pairing mechanism is likely phonon-mediated. Based on these observations, we assume that the leading superconducting instability is driven by an isotropic, short-range attractive interaction.
At the mean-field level, this leads to an $s$-wave–like pairing in momentum space, described by the following pairing Hamiltonian
\begin{equation}\label{eq:SI_M2BM_SCHam}
\begin{split}
\mathcal{H}_{\delta} =& \sum_{\mathbf{k}}\sum_{l}
\Delta_0
c_{-\mathbf{k},l,\downarrow}^\dagger c_{\mathbf{k},l,\uparrow}^\dagger
+\text{h.c.},
\end{split}
\end{equation}
where $\Delta_0$ denotes the uniform pairing amplitude.

In the weak pairing limit, where $|\Delta_0| \ll \alpha_R k_F$, 
inter-band pairing can be neglected. 
The effective superconducting pairing on the Fermi surfaces is then obtained by projecting the full pairing Hamiltonian $\mathcal{H}_\delta$ onto the low-energy subspace near the Fermi level.
Focusing on the $\epsilon_+(\mathbf{k})$ band, 
whose Fermi surface is spanned by the eigenstates $\chi_1$ and $\chi_4$, 
the projected gap function in this subspace takes the form
\begin{equation}\label{eq:SI_M2BM_SCProjFS14}
\begin{split}
\hat{\Delta}_{\text{eff}}
= \Delta_0 
\begin{pmatrix}
+\mathrm{i}
\exp(-\mathrm{i}\theta_{\mathbf{k}})
&0\\
0&
-\mathrm{i}
\exp(+\mathrm{i}\theta_{\mathbf{k}})
\end{pmatrix},
\end{split}
\end{equation}
which corresponds to a helical equal-spin pairing state.
It is straightforward to verify that the effective pairing described by Eq.~\eqref{eq:SI_M2BM_SCProjFS14} is fully gapped and preserves time-reversal symmetry, 
in full agreement with experimental observations.
Despite the spin splitting induced by spin-orbit coupling (SOC), 
the superconducting state remains robust and is not suppressed by SOC or inter-layer hybridization. 
This stability arises because Cooper pairs still form between time-reversal conjugate states,
consistent with the Anderson theorem.

This minimal model captures the essential features of systems with spin-orbit coupling induced by local inversion symmetry breaking, 
and it explains how effective helical equal-spin pairing can naturally arise in such systems. 
However, the Fermi surfaces in $\beta$-$\ch{Bi2Pd}$ exhibit a more complex topology, as all three $p$-orbitals of Bi contribute to the electronic structure near the Fermi level. 
A more detailed comparison with experiments therefore requires more sophisticated models informed by first-principles calculations.
Furthermore, due to the presence of both time-reversal and inversion symmetries, 
the superconducting state of $\beta$-$\ch{Bi2Pd}$ belongs to the symmetry class DIII, which supports topological phases characterized by a $\mathbb{Z}_2$ invariant in quasi-$2$D systems and a $\mathbb{Z}$ invariant in $3$D systems. 
Experimental evidence suggests the presence of topologically nontrivial states in the vortex cores~\cite{Lv2017}. 
Nevertheless, the precise nature of the topology is obscured by the complexity of the Fermi surfaces, 
and further investigation is needed to clarify the topological character of the superconducting state.

\end{document}